# Reconfigurable Compute-In-Memory on Field-Programmable Ferroelectric Diodes

Xiwen Liu,[a] John Ting,[a] Yunfei He,[a] Merrilyn Mercy Adzo Fiagbenu,[a] Jeffrey Zheng,[b] Dixiong Wang,[a] Jonathan Frost,[a] Pariasadat Musavigharavi,[a,b] Giovanni Esteves,[d] Kim Kisslinger,[e] Surendra B. Anantharaman,[a] Eric A. Stach,[b,c] Roy H. Olsson III,[a]* Deep Jariwala[a]*

[a] Electrical and Systems Engineering, University of Pennsylvania, Philadelphia, PA, 19104, USA

[b] Materials Science and Engineering, University of Pennsylvania, Philadelphia, PA, 19104, USA

[c] Laboratory for Research on the Structure of Matter, University of Pennsylvania, Philadelphia, PA, 19104, USA

[d] Microsystems Engineering, Science and Applications (MESA), Sandia National Laboratories, Albuquerque, New Mexico, 87185, USA

[e] Brookhaven National Laboratory, Center for Functional Nanomaterials, Upton, NY, 11973, USA

*Corresponding author: rolsson@seas.upenn.edu, dmj@seas.upenn.edu

**Abstract**

The deluge of sensors and data generating devices has driven a paradigm shift in modern computing from arithmetic-logic centric to data-centric processing. Data-centric processing require innovations at device level to enable novel compute-in-memory (CIM) operations. A key challenge in construction of CIM architectures is the conflicting trade-off between the performance and their flexibility for various essential data operations. Here, we present a transistor-free CIM architecture that permits storage, search and neural network operations on sub-50nm thick Aluminum Scandium Nitride ferroelectric diodes (FeDs). Our circuit designs and devices can be directly integrated on top of Silicon microprocessors in a scalable process. By leveraging the field-programmability, non-volatility and non-linearity of FeDs, search operations are demonstrated with a cell footprint < 0.12 $\mu m^2$ when projected onto 45-nm node technology. We further demonstrate neural network operations with 4-bit operation using FeDs. Our results highlight FeDs as candidates for efficient and multifunctional CIM platforms.

**Graphical Table of Contents**

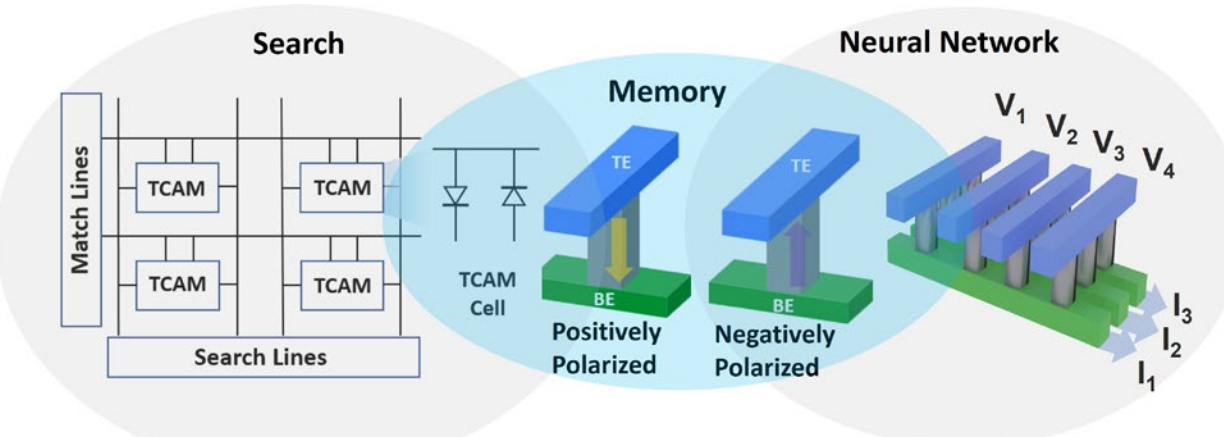

The convergence of big-data with artificial intelligence (AI) has led to multiple emerging technologies across a range of computing applications [1-3]. The increasingly ubiquitous presence of sensors and edge/IoT devices has created a flood of data, which has exposed a wide efficiency gap in computing hardware, ranging from mobile and edge devices to data centers [4-5]. In addition, the slowdown in the miniaturization of silicon-based complementary metal–oxide–semiconductor (CMOS) devices further accentuates the gap between resource requirements based on conventional von Neumann computing hardware architectures [6]. Furthermore, it is well-known that for many data-centric tasks in von Neumann architectures, most of the energy and time is consumed in memory access and data movement, rather than in actual computation [7-9]. Several solutions have been proposed to mitigate and overcome this bottleneck, with a prominent one being placing memory and logic units in close physical proximity. While significant progress has been made along those lines at both materials and device levels, a transformative approach would be to perform computing functions precisely where the data is stored using memory devices. This is popularly known as compute-in-memory (CIM). [8-9]. While several demonstrations of CIM architectures using novel non-volatile memories (NVMs) have been made, the bulk of the effort has been constrained to a single type of computing task, an example being matrix multiplication accelerators, typically achieved using memristive crossbar arrays [10-13]. However, AI computational tasks that exploit 'big-data' generally require more than one data-intensive computational operation on the same chip, preferably using the same architecture to process information in the pipeline. Three of the most important functions or operations are: 1) on-chip storage, 2) parallel search, and 3) matrix multiplication. A key challenge in the construction of CIM architectures is the conflicting trade-off between the performance and the flexibility required to achieve these three functions. Consequently, while CIM accelerators that can achieve high performance on matrix multiplication have been demonstrated, they are fundamentally ill-suited for other big-data operations such as parallel-search [11-13]. Hence, it is important to conceptualize and develop reconfigurable and operationally flexible hardware for CIM to simultaneously support essential data operations such as on-chip memory, parallel search and matrix multiplication acceleration [14].

In this work, we demonstrate circuit blocks based on FeD devices which support multiple, essential primitive data operations in-memory in a transistor-free design (Fig. 1), by leveraging the unique characteristics of aluminum scandium nitride (AlScN) ferroelectric diode (FeD) devices. First, we demonstrate CMOS back-end-of-line (BEOL) compatible FeD devices which are non-volatile and show self-rectifying behavior with a high non-linearity, a high ON/OFF ratio, and fast field-programming. Then, we exploit these unique properties and demonstrate a non-volatile ternary content addressable memory (TCAM) using cells comprising of 0-transistors/2-FeDs. This transistor-free approach is a key merit of our device and memory cell design. Consequently, the 2-FeD TCAMs have the most compact design known (0.12 $\mu m^2$/cell for 45 nm node) and are the first fully BEOL-compatible architecture in a transistor-free design. Finally, we also show that FeD devices can be programmed into 4-bit, distinct, conductive states with superior linearity and symmetry via electrical pulsing. Using this programmable, multi-bit attribute of the FeDs, we demonstrate a hardware implementation of neural network computation in the form of analog voltage-amplitude matrix multiplication. We demonstrate accuracies approaching ideal, software-based, neural networks. Our results indicate that the CMOS-compatible, field-programmable, non-volatile FeDs offer unique opportunities to build reconfigurable CIM architectures with superior balance between performance and flexibility.

Our FeD devices consist of a 45-nm thick layer of a sputter-deposited, ferroelectric AlScN layer sandwiched between top and bottom aluminum electrodes. This forms a metal-insulator-metal (MIM) structure, as shown in the left panel of Fig. 2a. AlScN is a recently discovered ferroelectric material with nearly ideal ferroelectric hysteresis loops, record values of remnant polarization, and a composition-tunable coercive field [15-17]. Furthermore, it can be integrated directly in a CMOS BEOL-compatible process technology over 8-inch wafers. It has also been shown to be one of the most promising candidates for high-performance ferroelectric memory devices, scalable down to < 10 nm in thickness [18]. The AlScN films were characterized electrically and exhibit large coercive fields, $E_C$, of 2-4.5 MV/cm [15-19]. This is important for scaling to thinner ferroelectric layers, all while maintaining a large memory window, high ON/OFF ratio and good retention [16,19]. When combined with high measured remnant polarizations – $P_r$ of 80-150 $\mu C/cm^2$ [15-19] – this leads to a significant tunneling electro-resistance effect, based on strong tunnel barrier modulation, thus giving a high ON/OFF ratio (See Experimental Methods and Supplementary Note 1 in Supporting Information). Figure 2(a) presents a representative cross-section transmission electron microscopy (TEM) image of the MIM FeD device comprised of an AlScN film with an Al top electrode, deposited on Al/AlScN/Si substrate. An atomic-resolution TEM image of the AlScN film is shown in Figure 2(b1).

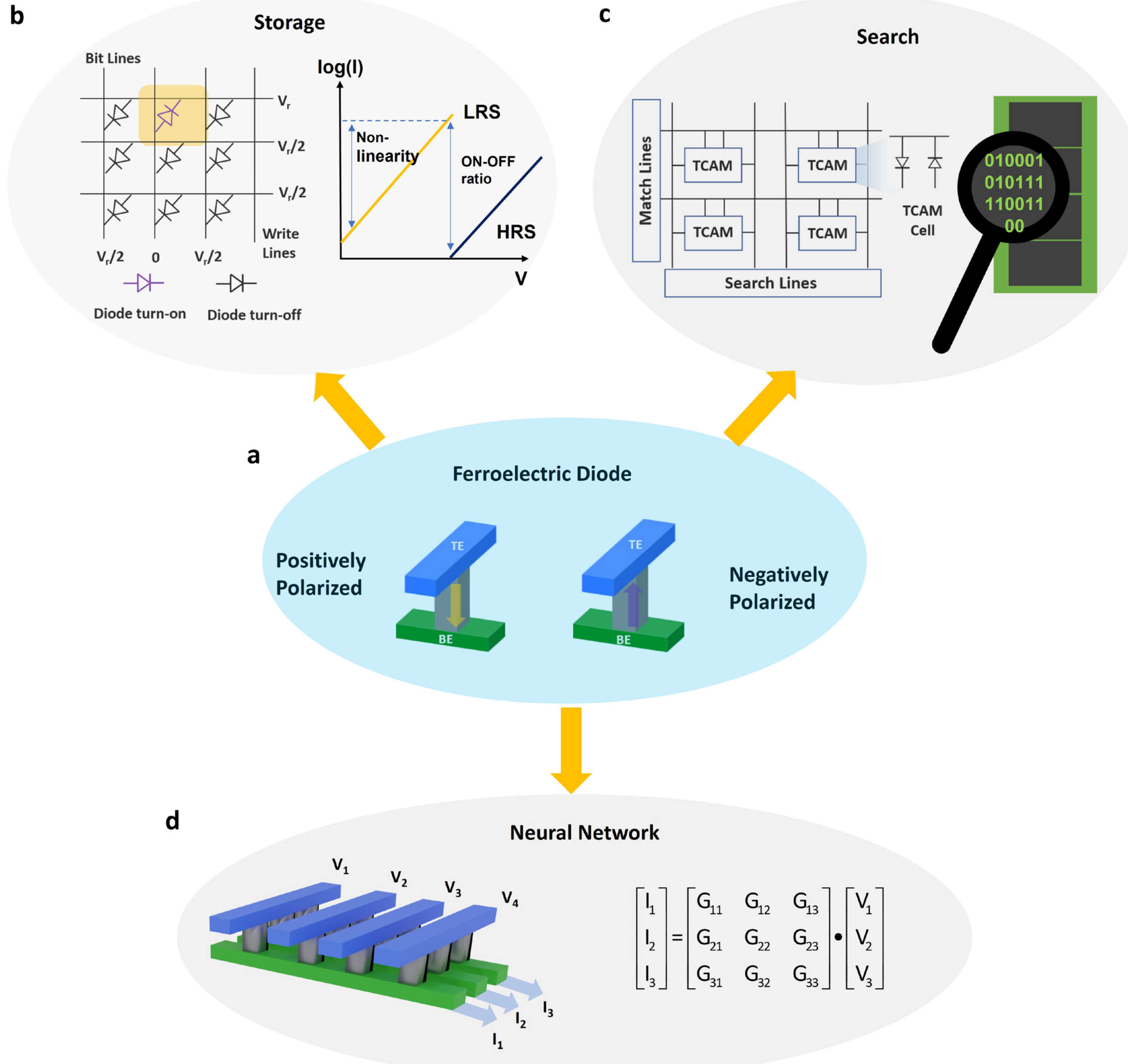

**Fig. 1. Reconfigurable CIM on Field-Programmable Ferroelectric Diodes.** a**,** Schematic diagram of FeD devices in a cross-bar structure with up and down polarization of the ferroelectric AlScN. The field programmability, non-volatility and non-linearity of these devices can be leveraged for multiple, primitive data operations such as storage, search, and neural networks without the need for additional transistors, as shown in b-d. b. The two-terminal FeD devices show a diode-like self-rectifying behavior with non-linearity > $10^6$ concurrently with a ON/OFF ratio over $10^2$ and write-endurance over $10^4$ cycles, making FeD devices well placed in the memory hierarchy for storage. Due to the high coercive field of AlScN, the read-disturbance would also be minimized which will result in high read-endurance. In addition, the high non-linearity can suppress sneak currents without the need for additional access transistors or selectors. C. For search operations, a non-volatile TCAM can be built upon 0-transistor/2-FeD cells, which serves as a building block

in hardware implementation of in-memory computing for parallel search in big data applications. D. For neural networks, FeD devices can provide programmability to distinct multiple conductive states with a high degree of linearity with respect to number of electrical pulses. This allows mapping the matrix multiplication operation, a key kernel in neural-network computation, into reading the accumulated currents at each bitline of a FeD device by encoding an input vector into analog voltage amplitudes and the matrix elements into conductances of an array of FeD devices. The matrix multiplication operation is benchmarked by mapping neural network weights to experimental FeD conductance states in a convolutional neural network architecture for both inference and the in-situ learning task, and shows that our accuracies approach ideal software-level simulation on the MNIST dataset.

The ferroelectric response of the 45 nm AlScN thin film was characterized by a positive-up, negative-down (PUND) measurement on a circular metal/ferroelectric/metal capacitor with radius of 25 µm, using a square wave with 2 µs delay and pulse widths of 400 ns (Supplementary Fig. S11). PUND testing is preferred over a polarization-electric field hysteresis loop (P-E loop) measurement (see Supplementary Fig. S16) because the P-E loop of 45 nm thick AlScN shows a polarization-dependent leakage which hinders the observation of polarization saturation for positive applied fields that switch the material into the metal-polar state [16-17]. The PUND result indicates a remnant polarization ~120 $\mu C/cm^2$, as shown in Fig. 2c, and in agreement with prior observations [15-19]. To further verify the ferroelectric switching, a dynamic current response was carried out, in which peaks corresponding to ferroelectric switching were observed (Supplementary Fig. S12). To further characterize the memory effect and reliability, we performed write endurance tests between the positive and negative polarization states, as shown in Fig. 2d. Fig. 2d presents remanent positive and negative polarization extracted from 20,000 PUND cycles. Cyclic set/reset operations of the same AlScN FeD device indicate that both positive and negative polarization states are stable and rewritable for a significant number of cycles. As shown in Fig. 2e, we repeatedly set/reset the FeD device between the low resistance state (LRS) and high resistance state (HRS) by applying negative/positive voltages on the top electrode while grounding the bottom electrode, for 100 cycles, using quasi-DC voltage sweeps (single I-V sweep shows a threshold-type resistive switching in Supplementary Fig. S15). The FeD device shows ultralow operating current and self-rectifying behavior with non-linearity > $10^6$ between 9 V and 0 V, which helps suppress sneak currents without the need for additional access transistors or selectors. A Poole-Frenkel tunneling model shows a good fit to experimental current-voltage data when compared to alternative tunneling mechanisms (Supplementary Fig. S17) and further discussion on the charge transport mechanism in the FeDs can be found in our previous work [19]. The distributions of LRS and HRS resistances are summarized in Fig. 2f showing a tight distribution on the cycle-to-cycle variation of the ratio between the LRS and HRS.

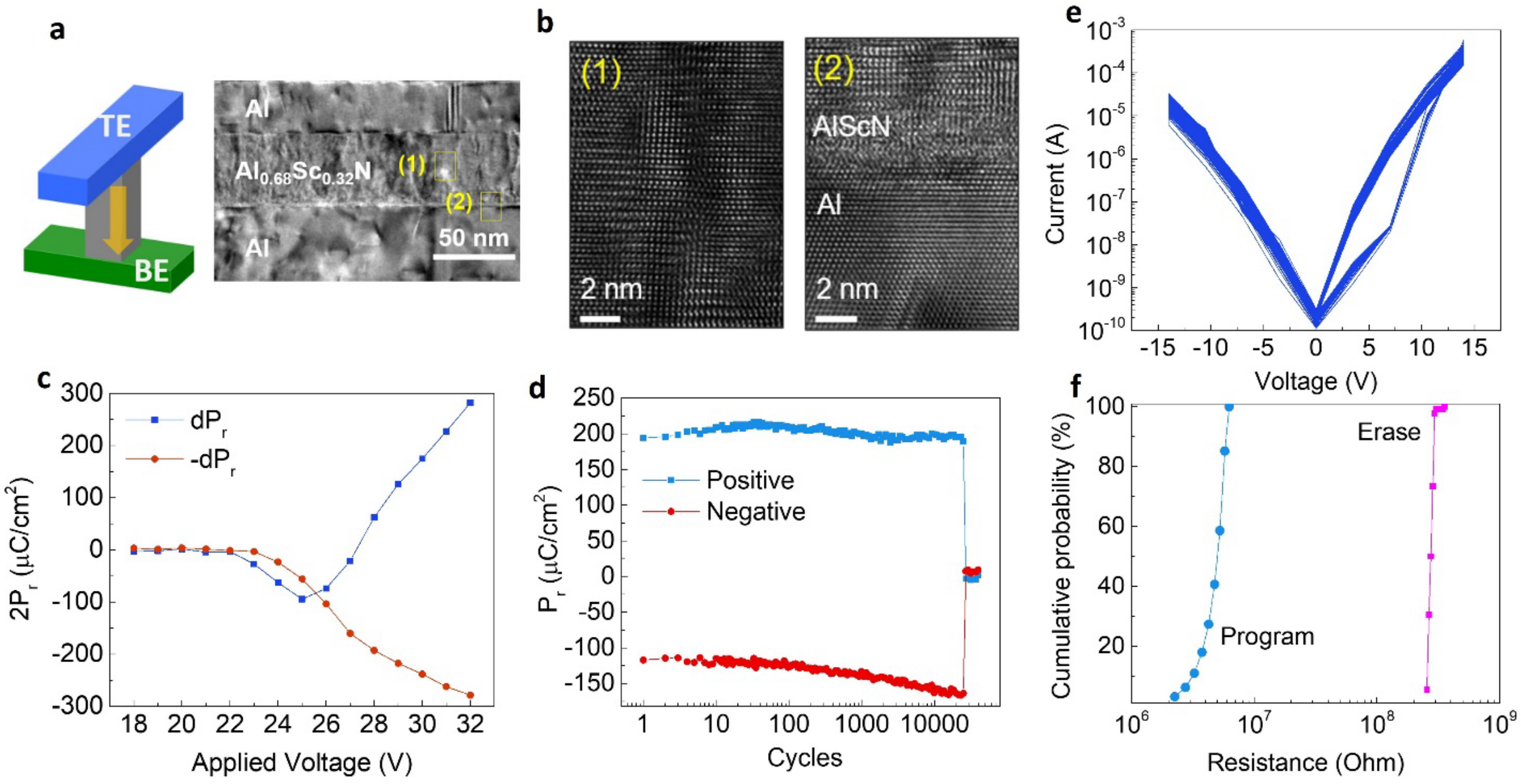

**Fig. 2. Room-temperature electrical characterization of AlScN FeDs**. a, 3D schematic illustration of the AlScN FeD device and cross-sectional TEM image of the AlScN FeD, showing 45 nm AlScN as the ferroelectric switching layer. b, The high-resolution phase-contrast TEM image obtained from the denoted regions (1) and (2) in (a) where the atomic structure of the ferroelectric and interface are visible. c, PUND results of a 45 nm AlScN thin film with a pulse width of 400 ns and 2 µs delay between pulses. The PUND test reveals a saturated remanent polarization of 150 µC/cm$^2$.d, The extracted remanent polarizations from PUND measurements during the endurance test of AlScN films using 1.5 µs pulse width and 26 V amplitude. e, 100 cycles of program and erase measurements over the 45 nm AlScN-based FeDs. f, Distribution of HRS and LRS resistances during program and erase measurements in e.

Next, we focus on CIM circuit architectures and computing applications comprising the above-described FeDs serving as non-volatile memories. TCAM is a key building block in hardware implementation of CIM for fast and energy-efficient parallel searches in big data applications [20-22]. A TCAM cell can be read in a search operation with an additional 'X' ('don't care') bit besides a bit value of either '0' or '1', which returns a match result regardless of the input search, making TCAM much more powerful and more universal in parallel searching applications. However, in a conventional Si CMOS architecture, multiple transistors (~16) are required to construct a single TCAM cell with static random-access memories (SRAMs) (Fig. 3a). This configuration results in large footprints and high-power consumption due to charging and discharging of the transistor and due to interconnect parasitic capacitances [21]. Non-volatile memories (NVMs) are promising alternatives for implementing TCAMs as they are more area-and energy-efficient [22]. TCAMs based on various NVMs have already been demonstrated [22-26]. However, all of those architectures are still constructed on top of front-end-of-line transistors and none of them are fully BEOL compatible.

In this work, the cell structure of TCAM can be significantly simplified by using just two FeDs (Fig. 3a). As shown in Fig. 3b, we write the logic ‘1’ state into the FeD TCAM cell by programming the left/right FeD to a low-resistance/high-resistance state, respectively. During the search operation, the match lines (MLs) are biased by a read voltage $V_S$ which is higher than the turn-on voltage of the FeD. Next, logic ‘1’ is searched by applying a high/low voltage to the search lines (SL and $\overline{SL}$), respectively, and vice versa for logic ‘0’. In this context ‘high voltage’ refers to the read voltage $V_S$ (8 V), which is higher than the turn-on voltage of the FeD, but is lower than the write voltage (15 V). Conversely, a ‘low voltage’ refers to a read voltage near zero. Since the left FeD is in parallel with the right FeD, a match state is observed only if both of the FeDs in a cell are cut-off (Fig. 3b, left panel). When the stored data and search data match (as shown in the left panel in Fig. 3b, the stored bit is logic ‘1’ and the search bit is logic ‘1’), the FeD with the low-resistance state is turned-off, as the voltage drop between its anode and cathode is near zero and lower than its turn-on voltage. Further, the FeD with the high-resistance state is also cut-off, because the current is naturally low when passing through the FeD in the high-resistance state. Therefore, the discharge currents at the two channels are both minimal and the ML stays high. However, when the search data does not match the stored data, the left FeD with the low-resistance state is turned-on as the voltage drop between its anode and cathode is high. Therefore, the discharge current is significant and the ML voltage is low (Figure 3b, middle panel). We also demonstrate a ternary ‘don’t care’ state in the two FeD-based TCAM by setting the left/right FeD both to a high-resistance state (Figure 3b, right panel). Whatever signals arrive at the two FeDs, both the FeDs are always cut off as they are in the high-resistance state. Fig. 3c shows repeated quasi-DC measurement of the resistance of the two FeD-based TCAM cell for both match and mismatch states between the search data and the stored data bit ‘1’, using search voltages of 7 V on the FeDs. Fig. 3d shows repeated quasi-DC measurement of the resistance of the two FeD TCAM cell for the stored data bit ‘Don’t care’, using both query bit ‘1’ and ‘0’. This shows that for both queries either “1” or “0”, the ML resistance of the two FeDs-based TCAM remains high and thus no discharging occurs through any of the two FeDs. Hence, the TCAM cell with two FeDs is fully functional with all three states.

Conventional two-terminal NVMs are always paired together with a front-end transistor to construct TCAM cells. This is because transistors are required to cut-off the channel, as they are in series with the two-terminal NVMs. The FeD-based design benefits from a high self-rectifying ratio which cuts off the channel without the need for any transistors. In other words, the FeD abstracts the functionality of the transistors into its self-rectifying behavior. The absence of a transistor leads to smaller cell footprints and superior area efficiency, and increases the search speed of the FeD-based TCAM. The equivalent resistor-capacitor (RC) circuits have been shown in Supplementary Fig. 14 for 2-FeD and 2T-2R based TCAMs, which could explain why the search delay in our FeD-based TCAM can be reduced in comparison to prior reported 2T-2R architecture. A benchmark comparison chart of lateral footprint of various TCAM cells is shown in Figure 3e.

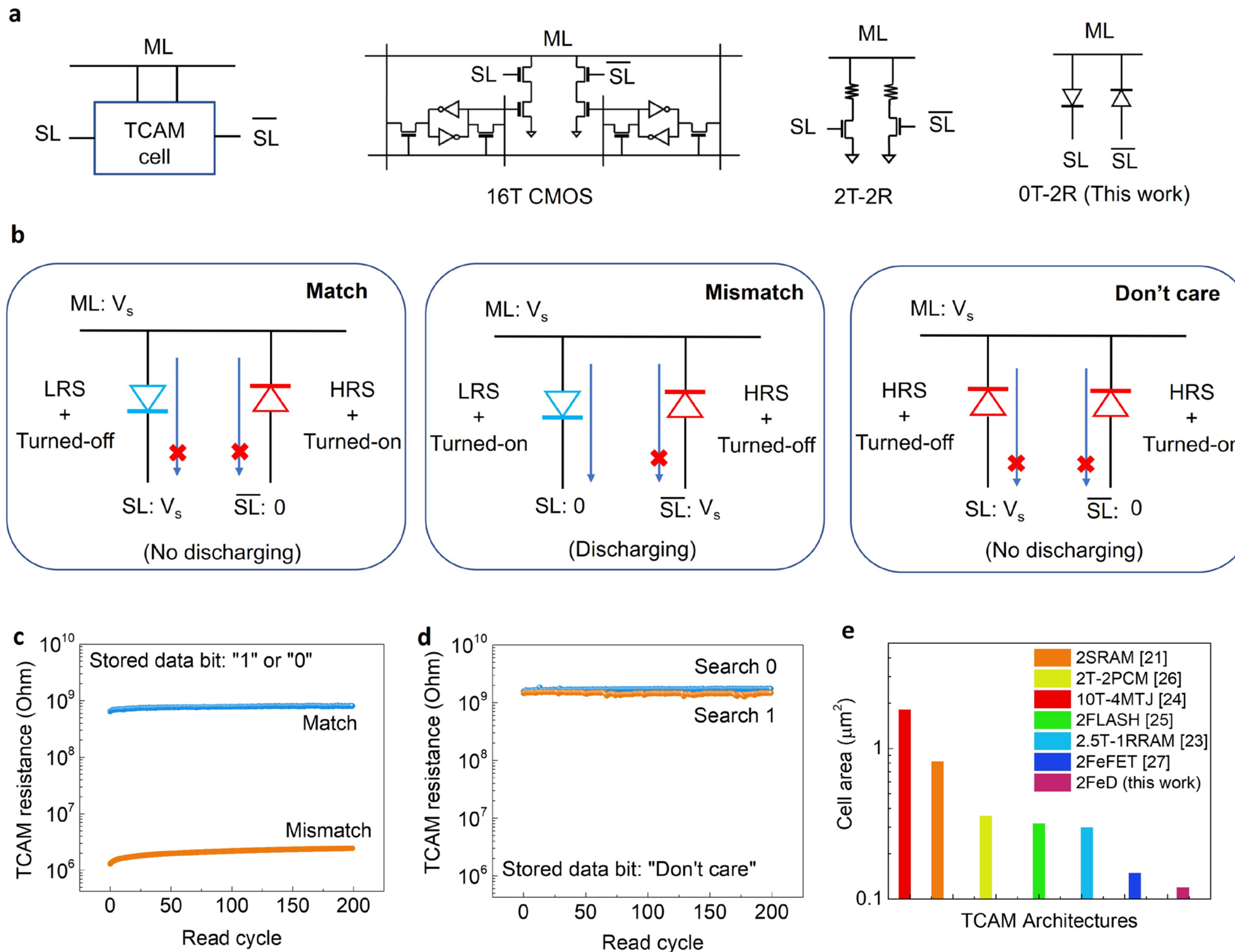

**Fig. 3. 2-FeD TCAM cell for search operation** a, A box schematic representation of a TCAM cell with match line (ML), search line (SL) and search line bar (SL bar) electrodes (left). Circuit diagrams of a single 16 transistor (16T) TCAM cell based on CMOS SRAM technology, and 2-transistor-2-resistor (2T2R) TCAMs based on resistive storage elements such as PCM and RRAM (center). The two FeD-based TCAM cell proposed in this work (right) significantly simplifies the TCAM design by using two FeDs connected in parallel but oppositely polarized. The cell structure makes it natural to utilize the FeD in crossbar memory arrays, in which the signal lines connecting to the anode and to the cathode are in parallel in a bit-search for the TCAM demonstration, as shown in Supplementary Fig. S3. b, Operation of a single TCAM cell comprising 2 FeDs for “match”, “mismatch” and “don’t care” states. The full look up table of the two FeD-based TCAM cell is summarized in Supplementary Table. 1. c, Repeated, quasi-DC extraction of the resistance

of the two FeDs TCAM cell for both match and mismatch states between the search data and the stored data bit '1', showing a >100 X difference over ML resistances. d, Repeated quasi-DC measurement of the resistance of the two ferro-diodes TCAM cell for the stored data bit 'Don't care', using both query bit '1' and '0', which turns out that for both two queries the ML resistance of the two FeDs TCAM is high and thus no discharging through any of the two FeDs occurs. The sensing margin of our FeDs-based TCAM is a function of both the self-rectifying ratio and the ON/OFF conductance (or current) ratios [26]. Per our detailed compact model (See supplementary note 1) the ON/OFF ratio of a FeD can be further improved by integrating a non-ferroelectric insulator on top of the ferroelectric layer and engineering both the thickness ratio between these ferroelectric and non-ferroelectric insulator layers as well as the coercive field of the ferroelectric layer. Future studies will focus on further improving the sense margins by engineering these variables. e, A benchmark comparison chart of lateral footprint of TCAM cells in various memory technologies [21, 24-27]: resistive random-access memories (RRAMs) [22-23], magnetic tunnel junction (MTJ) RAMs [24], floating gate transistor memory (FLASH) [25], phase change memories (PCMs) [26], and ferroelectric field-effect-transistor (FeFET) [27]. A single FeD area of 0.0081 $\mu m^2$ is obtained for this estimate. The superior performance of our two FeD-based TCAM when compared to CMOS SRAM and other transistor plus-NVM devices-based architectures is evident.

Next, we focus on the application of our FeD device arrays for deep neural network (DNN) inference/training, which involves repeated matrix multiply/accumulate (MMAC) operations. [8, 11-13]. The ideal MMAC-suitable memristive devices should perform linearly arranged conductance values over electrical programming, excellent ohmic behavior, and high resistance to suppress the amount of current. Prior research has primarily focused on memristive devices which exhibit excellent ohmic behavior and a large number of conductance states, such as RRAM [11] and PCM [29]. In the context of DNN inference accuracy, an excellent ohmic behavior is necessary to minimize the distortion of the input datum and a large number of conductance values will minimize the precision loss on the weight matrix. However, those requirements will hurt the architecture metrics for power efficiency and low latency per computation. First, memristive devices with excellent ohmic behavior increase the amount of operating current and further impose limitations on array scaling-up [11, 13, 29]. Second, a large number of conductance values will correspondingly require high-precision analog-to-digital converters (ADCs) [30-32]. It is already known that the energy and area costs are dominated by the ADCs in memristor array systems [30-32]. Therefore, more conductance states mean more power overhead at the architecture level. Here we show that FeD memristors can be used to perform an optimal trade-off between these metrics. First, to relax the trade-off on the device conductance, it is important to decrease the operating conductance of memristive devices while maintaining linear behavior. The former condition is readily met for highly self-rectifying devices, an inherent property of the FeDs; the latter condition can be met by applying an encoder on the input voltage amplitudes to linearize the current-voltage relationship (See Supplementary Note 2 in Supporting Information) [13]. Second, to relax the trade-off on the number of conductance states, fewer but sparsely and linearly arranged conductance states are necessary. This approach can achieve equivalent inference accuracy in comparison to approaches implementing a large number of conductance states [33-34].

Fig. 4a shows the gradual switching in a FeD by stepwise voltage pulse modulation (ranging from 16 V to 19 V). The FeD cells are gradually programmed into 16 distinct conductance states with a high degree of linearity using stepwise voltage pulses, followed each time by an erase operation. Fig.4b shows that the FeD device is capable of voltage pulse-induced analog bipolar switching (ranging from 16 V to 19 V). The FeD device showed superior linearity ($R^2$ score of 0.9997 for a linear fit) for 16 distinct conductance states over bidirectional modulation. Conductance retentions for 16 distinct conductance states are shown in Fig. 4c, and show no obvious degradation. Fig.4d shows conductance states distribution of five separate FeD devices subjected to the identical sequences of 16 program pulses (2 μs pulse width) with interleaved reads (8 V). The results show negligible device-to-device variations between those FeD devices. We evaluate the performance of arrays comprising FeD devices for neural network and MMAC applications in two situations: 1) inference phase post training; 2) in-situ training. For the inference applications, a convolutional neural network (CNN) (as shown in Fig. 4e) was first trained using single-precision floating-point format on the MNIST dataset [35], after which the weights in the CNN architecture are extracted and stored. The pretrained weights in the CNN are then quantized layer-wise by normalizing the weights to the FeD conductance range as shown in Fig. 4e. An added varying factor A, which is an indicator of non-linearity together (see Supplementary Note 3 in Supporting Information for details) with low-precision quantization, have been modeled in the weight mapping process. To quantify the effects of nonlinearity and asymmetry on the accuracy loss, the pre-trained weights of the CNN are normalized to a number of conductance states (from 9 to 1-bit) with various A factors (from 0.1 to ideal) before the inference phase. No significant accuracy loss in the CNN has been observed for low-precision weight quantization with low non-linearity ($A > 0.5$), as can be seen in Fig. 4f. The benchmarked software testing accuracy of 97.5% on floating-point format is sustained with just 3-bit (and more) effective weight per FeD device if weight non-linearity is highly suppressed ($A > 0.35$). High non-linearity ($A < 0.35$) requires one- or two-bits weight precision to perform equivalent testing accuracy on floating-point format, which indicates that sparsely arranged conductance states with superior linearity can replace a large amount of conductance states for equivalent inference accuracy. Furthermore, we simulate the in-memory implementations of in-situ training on FeD arrays where the same CNN is being trained and weight updates are directly mapped to the realistic conductance states of the demonstrated FeDs after each backpropagation. The high linearity of weight distribution in the training phase is more important than inference as the non-linearity and distortion of the weights could lead to opposite gradients obtained through backpropagation and hurt the neural network training. As shown in Fig. 4g, for the demonstrated 16 separate conductance states in Fig. 4a in the FeD devices, the in-situ learning accuracy suffers a ~2% degradation compared to the accuracy trained by software on floating-point number.

The above experimental demonstration of a TCAM memory cell and neural network have been performed on FeD devices with 45 nm thick FE AlScN layer. This means the operating voltages are far above current high node Si CMOS operating voltages. However, we also note that FE-AlScN FeDs have already been scaled to 20 nm thick devices operating in the 5 V range [19]. Further our group has also demonstrated FeDs from 100 nm thick AlScN and the voltage scaling with reducing thickness is well aligned with thickness scaling (See Figure S18 for details). Given that 9 nm thick FE AlScN films with > 100 μC/cm$^2$ $P_r$ have been demonstrated [18] we expect that our TCAM cell circuit and neural network concept can be extended to be voltage compatible with higher end Si CMOS nodes at < 2 V operation. Further, it is worth noting that while the write voltages in our current experimental demonstration of TCAM are high, the read voltages are lower by ~2x. Therefore, in an application such as a TCAM it may be acceptable to have a higher and less efficient write-voltages if the read-operations are low-voltage and low-power. None the less, full circuit level simulations using SPICE of TCAM cells of scaled FeD devices do suggest that the search delay for our architecture is in the 100 ps range (see Figure S19 for more details) presenting an advantage in terms of low latency of search at high densities when compared to competing technologies.

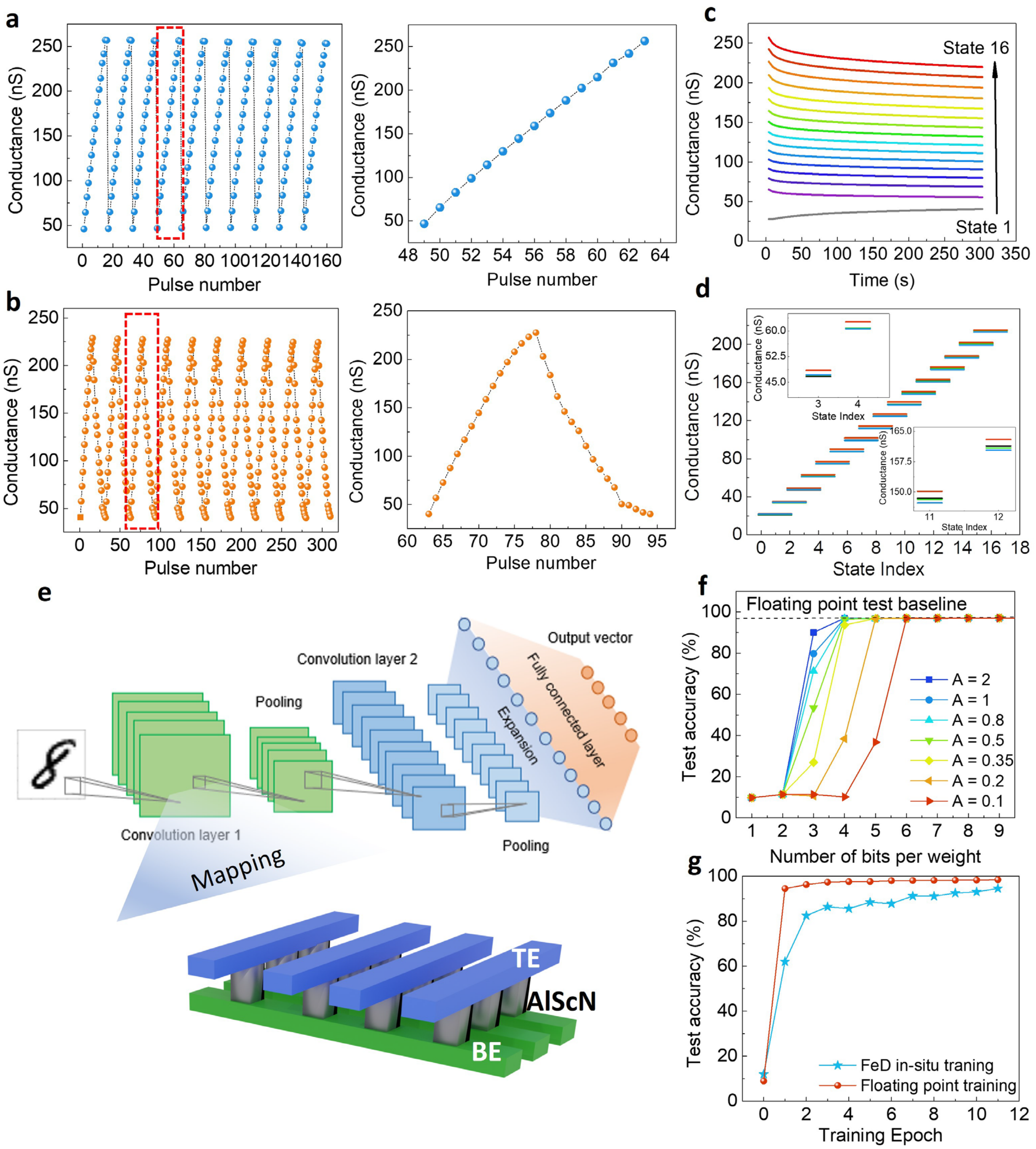

**Fig. 4. FeD-based neural network.** a, Gradual switching in FeD device by stepwise voltage modulation pulses. The FeD device showed superior linearity over 16 distinct states. Callout window (right panel) shows one cycle of gradual programming followed by an erase operation. b, The FeD is demonstrated to be capable of voltage pulse-induced analog bipolar switching (left). Callout window (right) shows one cycle of gradual programming and gradual erasing. We note that the range of conductance in FeD devices used to program these states (~25-250 nS) is much

smaller as compared to the range of conductance used for TCAM operations ( ~2-250 nS). This is primarily because this linearity in operation is better achieved in a smaller range of conductance. Further, the DNN inference application does not necessarily require a high range of conductance modulation [10-11, 34]. c, Resistance retentions for 16 distinct resistance states. d, Resistance states distribution of five separate FeDs subjected to sequences of 16 program pulses (2 µs pulse width) with interleaved reads (8 V). e, Illustration of a CNN (including two convolutional layer and one fully connected layer) trained and weight mapping for the MNIST dataset. As the hardware implementation of a neural network using FeD arrays can operate in a fully analog domain, the peripheral analog-to-digital converters could be waived. f, Simulated performance of inference efficacy of the network in (e). In inference phase predictions are made by the quantized CNN over the MNIST test dataset and inference accuracy is extracted. The simulations in (f) demonstrate that degradation of the network's inference accuracy is less than 1% for low weight precision of just 3 bits for $A < 0.5$. g. Simulations of in-situ training of the network in (e) directly with the FeD devices implementing analog weight layers. Leveraged by the superior linearity in the gradual programming of the FeDs, the analog weight layers with 16 resistance states have been simulated to perform at a training accuracy comparable to the software baseline on floating-point numbers. This accuracy loss can be further reduced with more advanced low-precision training techniques and model compression techniques on software [33].

In conclusion, we demonstrate AlScN based FeD devices as a novel, BEOL compatible platform for multi-functional CIM in a transistor-free architecture. Our experimental demonstration of a search function is realized via a TCAM circuit with a lateral cell footprint bettering all existing and experimental NVM technologies. Finally, we demonstrate a stable, pulse-programmable 4-bit memory from FeD devices combined with a hardware implementation of a convolutional neural network with inference accuracy comparable to software. Our work therefore opens new possibilities in CIM platforms by enabling architectures with novel ferroelectric materials and diode devices made using them.

## ASSOCIATED CONTENT

### Supporting Information

The supporting information contains experimental methods of materials growth, device fabrication, device characterization as well as additional experimental details on electrical characterization and analysis of the ferroelectric diode charge transport, simulation results. The supporting information also contains details of the device compact model to evaluate the performance as well as TCAM circuit simulations, diagrams and comparative analysis.

The Supporting Information is available free of charge on the journal website at DOI:

AUTHOR INFORMATION

**Corresponding Author**

*E-mail: rolsson@seas.upenn.edu , dmj@seas.upenn.edu

**Author Contributions**

X.L., D.J., R.O. and E.S. conceived the idea. X.L. designed TCAM circuits and the reconfigurable CIM architecture. X.L. performed all neural network simulations and SPICE simulations. X.L., J.T., Y.H., M.M.A.F., S.B.A performed all the electrical measurements. X.L., J.F. and Y.H. performed all compact modeling. D.W., J.Z., and G.E. performed microfabrication, ferroelectric material and metal thin film growth. P.M. and K.K. performed transmission electron microscopy imaging (P.M.) and sample preparation (K.K). D.J., R.O., E.S. and X.L. analyzed and interpreted the data. All others contributed to writing of the manuscript.

**Notes**

**Competing interests**

The authors declare the following competing interests: D.J., X.L., R.O. and E.A.S. have a provisional patent filed based on this work. The authors declare no other competing interests.

**Data Availability**

The data that support the conclusions of this study are available from the corresponding authors upon reasonable request.

**ACKNOWLEDGEMENTS**

This work was supported by the Defense Advanced Research Projects Agency (DARPA), Tunable Ferroelectric Nitrides (TUFEN) Program under Grant HR 00112090047. The work was carried out in part at the Singh Center for Nanotechnology at the University of Pennsylvania which is supported by the National Science Foundation (NSF) National Nanotechnology Coordinated Infrastructure Program (NSF grant NNCI-1542153). The authors gratefully acknowledge use of facilities and instrumentation supported by NSF through the University of Pennsylvania Materials Research Science and Engineering Center (MRSEC) (DMR-1720530). D.J. acknowledges support from the Intel RSA program. This research used resources of the Center for Functional Nanomaterials, which is a US Department of Energy Office of Science User Facility, at Brookhaven

National Laboratory under contract no. DE-SC0012704. The data that support the conclusions of this study are available from the corresponding authors upon request.

Supporting information

# Reconfigurable Compute-In-Memory on Field-Programmable Ferroelectric Diodes

Xiwen Liu,[a] John Ting,[a] Yunfei He,[a] Merrilyn Mercy Adzo Fiagbenu,[a] Jeffrey Zheng,[b] Dixiong Wang,[a] Jonathan Frost,[a] Pariasadat Musavigharavi,[a,b] Giovanni Esteves,[d] Kim Kisslinger,[e] Surendra B. Anantharaman,[a] Eric A. Stach,[b,c] Roy H. Olsson III,[a]* Deep Jariwala[a]*

[a] Electrical and Systems Engineering, University of Pennsylvania, Philadelphia, PA, 19104, USA

[b] Materials Science and Engineering, University of Pennsylvania, Philadelphia, PA, 19104, USA

[c] Laboratory for Research on the Structure of Matter, University of Pennsylvania, Philadelphia, PA, 19104, USA

[d] Microsystems Engineering, Science and Applications (MESA), Sandia National Laboratories, Albuquerque, New Mexico, 87185, USA

[e] Brookhaven National Laboratory, Center for Functional Nanomaterials, Upton, NY, 11973, USA

*Corresponding author: rolsson@seas.upenn.edu, dmj@seas.upenn.edu

**Experimental Methods:**

**Device fabrication**

FeD consisted of a film stack of Al (80 nm)/$Al_{0.68}Sc_{0.32}N$ (45 nm)/Al (30 nm) on top of a Si/$Al_{0.8}Sc_{0.2}N$ (85 nm) substrate. To prepare this stack, we start by sputter depositing an 85-nm thick $Al_{0.80}Sc_{0.20}N$ template on the top of a 6” Si <100> wafer. The $Al_{0.8}Sc_{0.2}N$ was deposited using pulsed-DC reactive sputter deposition from a single-alloyed $Al_{0.8}Sc_{0.2}$ target using 5 kW of target power, a pressure of $7.47\times10^{-3}$ mbar, and a deposition temperature of 375°C. The first layer of 85 nm $Al_{0.8}Sc_{0.2}N$ serves to orient the subsequent 80-nm thick Al layer into a {111}-orientation. This Al (80 nm thick) layer serves as the bottom electrode for the second layer of $Al_{0.68}Sc_{0.32}N$ (45 nm thick), which is the ferroelectric layer used in this device. The 45-nm thick ferroelectric $Al_{0.68}Sc_{0.32}N$ film was co-sputtered from separate 4-inch Al and Sc targets in an Evatec CLUSTERLINE® 200 II pulsed DC Physical Vapor Deposition System. The Al and Sc targets were operated at 1250 W and 695 W, respectively, at a chuck temperature of 350 °C with 10 sccm of Ar gas flow and 25 sccm of $N_2$ gas flow. The chamber pressure was maintained ~$1.45\times10^{-3}$ mbar. This sputter condition resulted in a deposition rate of 0.3 nm/sec. The highly oriented {111} Al layer promotes the growth of AlScN with its [0001] axis direction being perpendicular to the substrate, thus, yielding a highly textured FE film. Then, without breaking vacuum, a 30 nm Al layer was sputtered as the top electrode and as a capping layer to prevent the oxidation of ferroelectric $Al_{0.68}Sc_{0.32}N$ surface.

**Device characterization**

Current-voltage measurements were performed in air at ambient temperature using a Keithley 4200A semiconductor characterization system. P-E hysteresis loops and PUND measurements of ferroelectric AlScN were conducted using Keithley 4200A semiconductor characterization system and a Radiant Precision Premier II testing platform. The TEM cross-sectional sample was prepared in a FEI Helios Nanolab 600 focused ion beam (FIB) system using the in-situ lift-out technique. The sample was coated with a thin carbonaceous protection layer by writing a line on the surface with a Sharpie® marker. Subsequent electron beam and ion beam deposition of Pt protection layers were used to prevent charging and heating effects during FIB milling. At the final cleaning stage, a low-energy Ga+ ion beam (5 keV) was used to reduce FIB-induced damage. TEM characterization and image acquisition were carried out on a JEOL F200 operated at 200 kV accelerating voltage. The sample was orientated to the [001] zone axis for imaging. All of the captured TEM images were collected using Digital Micrograph software.

**Supplementary Note 1** Compact model for evaluation of ON/OFF ratios of ferroelectric diode

1. General method of ON/OFF ratio evaluation

To evaluate the ON/OFF ratio and capture the IV characteristics of the ferroelectric diodes (FeDs), generally, we need to resolve the electron transport in the ferroelectric diode. Within the ferroelectric, there are three main methods of electron transport: direct tunneling, Fowler-Nordheim tunneling, and thermionic emission, the band diagrams illustrating each effect are shown in Fig.1. We can use the Wentzel-Kramers-Brillouin (WKB) approximation to encompass all three methods with one formula. In the approximation, the tunneling probability is given by:

$$T(E) = \exp(-\frac{2}{\hbar}\int_{x_1}^{x_2}\sqrt{2m^*[\left(E_c + qV(x)\right) - E]}\mathrm{d}x) \tag{1.1}$$

where $\mathrm{m}^*$ is the effective mass of the electron, $\mathrm{E_c}$ is the coercive field of the ferroelectric, $\mathrm{V(x)}$ is the voltage across the ferroelectric, and $\mathrm{E}$ is the applied field. We can give the integrand for the density of states as

$$N(E) = k_B T \ln\left(\frac{1+\exp\left(\frac{E-E_{f,1}}{k_B T}\right)}{1+\exp\left(\frac{E-E_{f,2}}{k_B T}\right)}\right) \tag{1.2}$$

where $\mathrm{K_B T}$ is the Boltzmann constant multiplied by temperature, $\mathrm{E_{f,1}}$ and $\mathrm{E_{f,2}}$ are the Fermi-levels at the left and right of the ferroelectric. With these formulas, we can define the current density J as

$$J = \frac{4\pi m^* q}{\hbar^3}\int_{E_{min}}^{E_{max}} T(E)N(E)\mathrm{d}E \tag{1.3}$$

This current density multiplied by the area of the ferroelectric film gives the tunneling current through the device. While this model does effectively capture the I-V characteristics of a ferroelectric diode, it lacks efficiency.

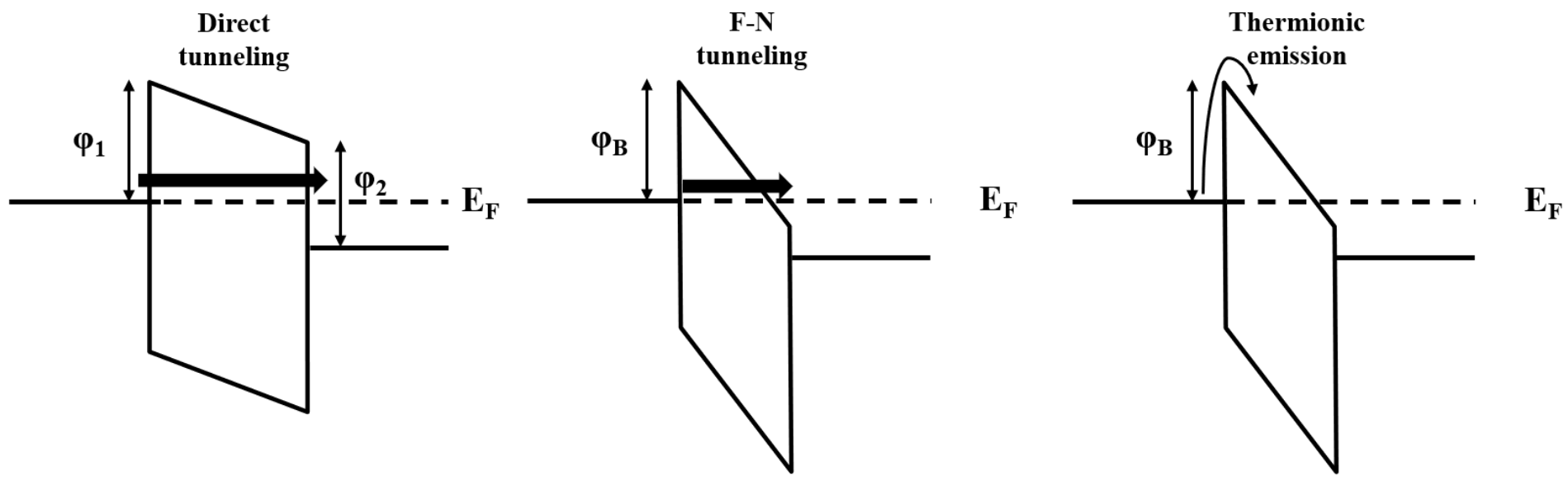

**Fig. S1.** Illustrations of the three methods of electron transport through a potential barrier. (a) direct tunneling (b) Fowler-Nordheim tunneling (c) thermionic emission.

## 2. Voltage shifting model and its verification.

### 2.1. Voltage shifting model.

In the analysis, we proposed a new compact shift model to describe the I-V characteristics of the FeDs. In the FeD, we can treat the changes in the I-V curve of the diode from the low resistance state (LRS) to the high resistance state (HRS) as the I-V curve shifting from left to right by the amount of $\Delta V$ as the diode transforms from LRS to HRS. In other words, it takes more voltage to offset the voltage of $\Delta V$ at HRS to generate the same current as the current at LRS. The shifted voltage $\Delta V$ can be derived as:

$$\Delta V = E_{dp}t \tag{2.1}$$

where t is the thickness of the ferroelectric layer, and $E_{dp}$ is the depolarization field of the ferroelectric, and we found that we can express the depolarization field as:

$$E_{dp} = E_C \tanh(\frac{\beta P_r}{\epsilon_{fe} E_C}) \tag{2.2}$$

where $E_C$ is the coercive field of the ferroelectric layer, $P_r$ is the remanent polarization, $\epsilon_{fe}$ is the dielectric constant of the ferroelectric, and $\beta$ is the parameter related to the oxide capacitance $C_{ox}$ and the ferroelectric capacitance $C_{fe}$, as [2]:

$$\beta = \left[(\frac{C_{ox}}{C_{fe}} + 1)\right]^{-1} \tag{2.3}$$

2.2. Comparison between the voltage shifting model and WKB approximation.

We verified that the voltage shifting model showed a relatively consistent result as the I-V characteristics of the ferroelectric diode modeled by the WKB approximation, shown in Fig. S2. In addition, the voltage shifting model is much more efficient since it utilizes analytical equations as opposed to numerical integration as the WKB approximation did. Furthermore, the new model's hyperbolic function helps us focus on the readout voltage region of the I-V curve since the depolarization field $\mathrm{E_{dp}}$ would not exceed the coercive field $\mathrm{E_c}$ in the model.

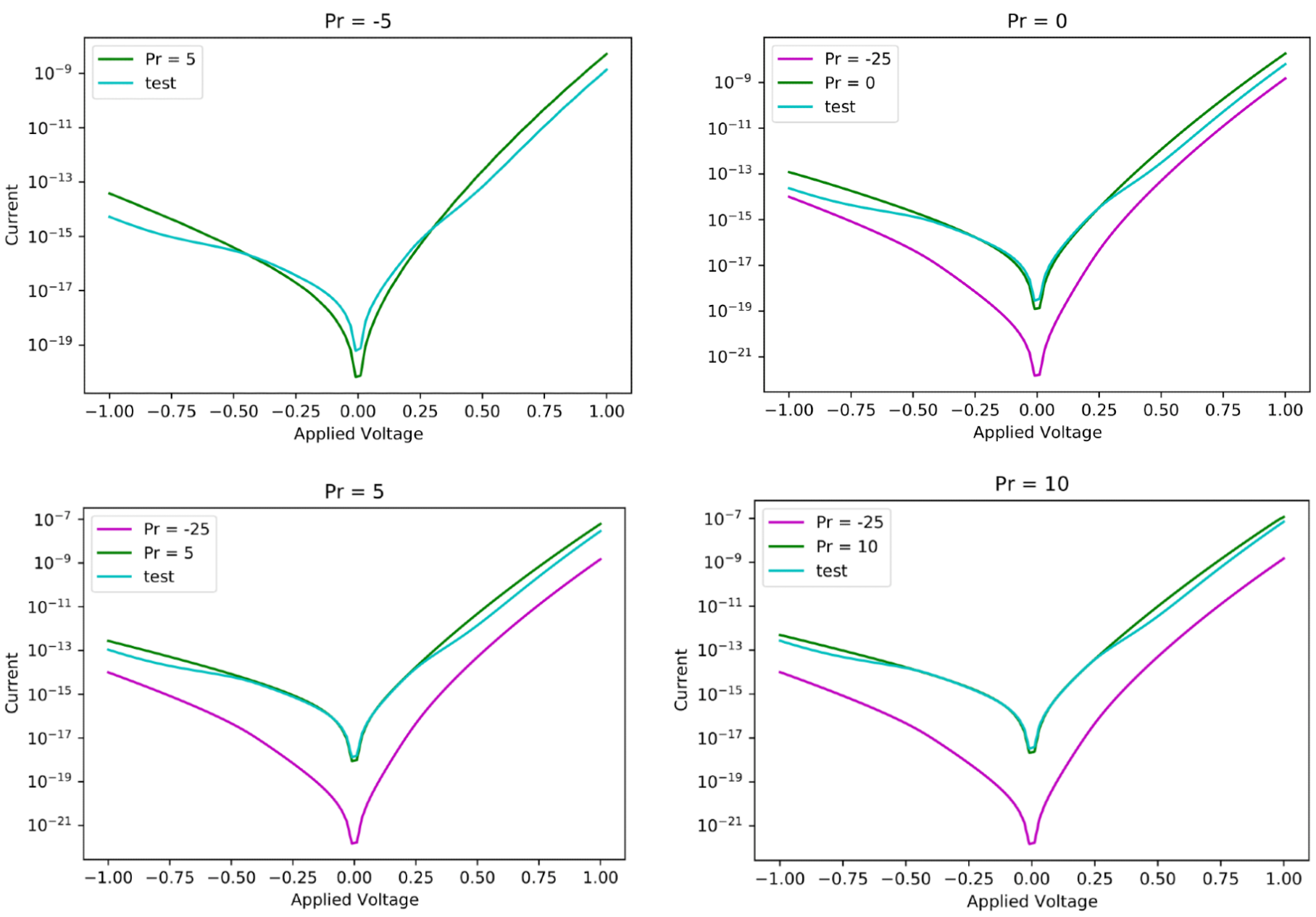

**Fig. S2.** Comparisons of the IV curves using the WKB approximation (green) vs. the new voltage shifting model for a ferroelectric diode (blue) for any arbitrary ferroelectric. The purple curve represents the base curve at high-resistance state modeled with the WKB approximation.

3. I-V Curves and ON/OFF ratio related to the voltage shifting model.

After the verification of this compact model, we used this model to find the overall trend of the relationships between the ON/OFF ratios of the ferroelectric diodes and parameters such as oxide capacitance, remanent polarization, and coercive field.

3.1. I-V curves and ON/OFF ratio changing with oxide capacitance.

First, we plot the I-V curve of the FeD shifting under different insulator capacitance. We pre-coded the oxide capacitance into $\beta$ using eq.2.3, and $\beta$ should vary between 0 and 1. By choosing the proper value of remanent polarization and coercive field, the simulated I-V curve of HRS shifting is shown in Fig. S3(a). Correspondingly, the ON/OFF ratio under different β is demonstrated in Fig. S3(b).

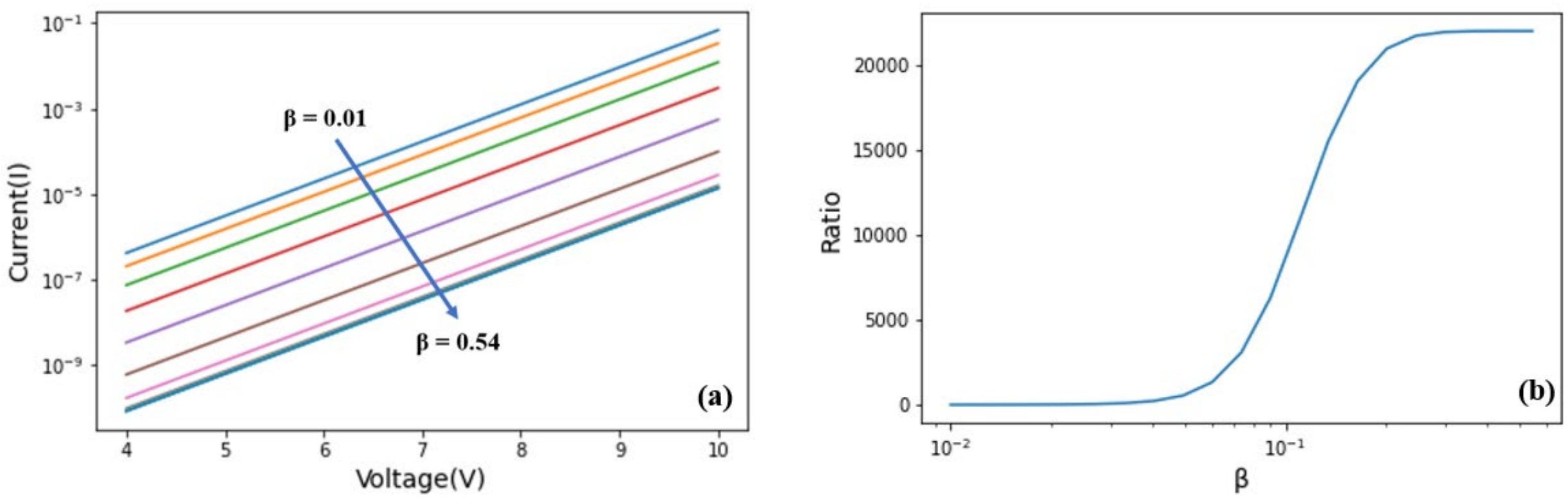

**Fig. S3.** The plot of HRS I-V curve shifting with the variation of the pre-coded oxide capacitance β is shown in (a), we vary β from 0.01 to 0.54 exponentially to get a better view of the I-V curve shifting trend. (b) is the ON/OFF ratio vs. β, and the ON/OFF ratio is the current ratio between LRS and corresponding shifted curve at the voltage of 7 V in (a).

3.2. I-V curves and ON/OFF ratio changing with remanent polarization.

Second, we vary the remanent polarization $\mathrm{P_r}$ from 1 to 135 $\mu\mathrm{C/cm^2}$ with a suitable insulator capacitance $\mathrm{C_{ox}}$ and coercive field $\mathrm{E_C}$, and the resulting I-V characteristics are shown in Fig. S4(a). The corresponding ON/OFF ratio under different remanent polarization is demonstrated in Fig. S4(b).

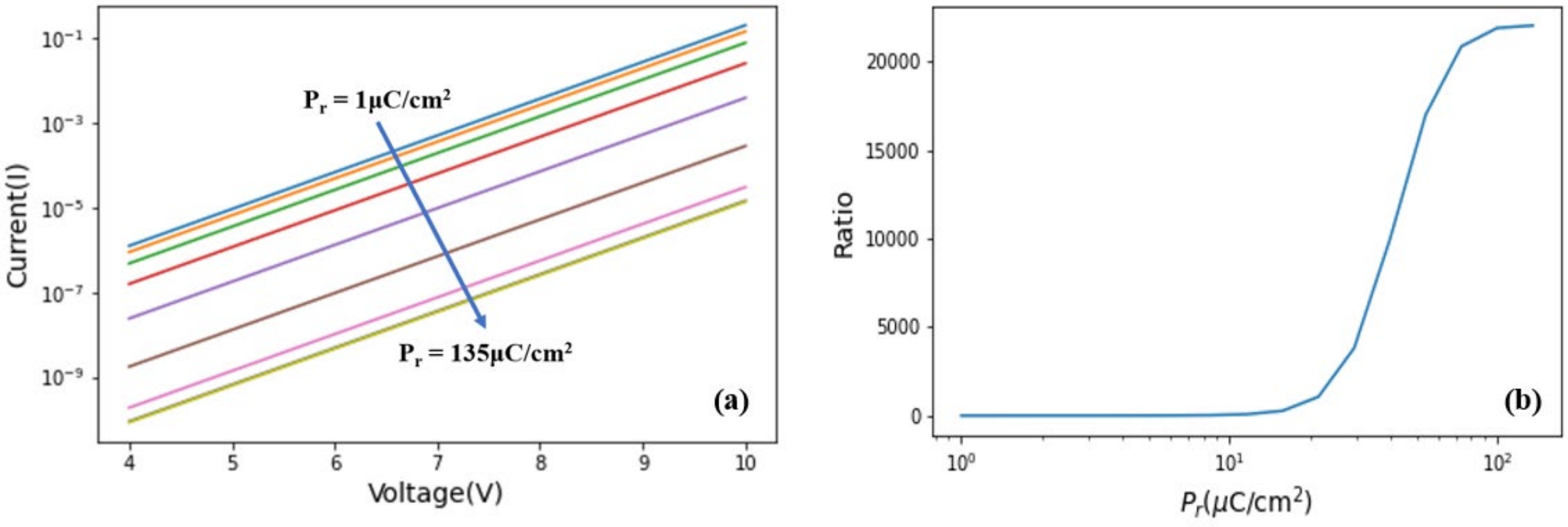

**Fig. S4.** The plot of HRS I-V curve shifting with the variation of the remanent polarization $\mathrm{P_r}$ is shown in (a), we vary $\mathrm{P_r}$ from 1 to 135 $\mu\mathrm{C/cm^2}$ exponentially to get a better view of the I-V curve shifting trend. (b) is the ON/OFF ratio vs. $\mathrm{P_r}$, and the ON/OFF ratio is the current.

### 3.3. I-V curves and ON/OFF ratio changing with coercive field.

Last, the I-V curve shifting plot for coercive fields $E_c$ linearly ranging from 0.12 to 3.12 $MV/cm$ is shown in Fig. S5(a). The corresponding plot of the ON/OFF ratio of the FeD under different coercive fields is shown in Fig. S5(b).

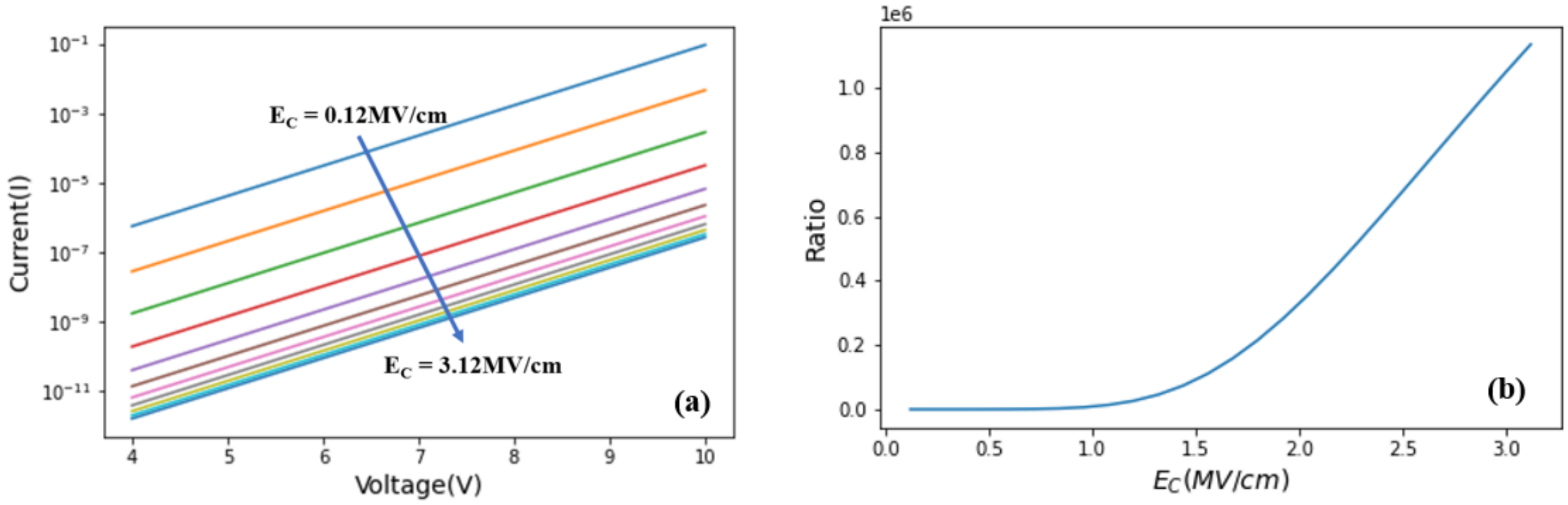

**Fig. S5.** The plot of HRS I-V curve shifting with the variation of the result of the coercive field $E_C$ is shown in (a), we vary $E_C$ from 0.12 to 3.12 $MV/cm$ linearly to get a I-V curve shifting trend. (b) is the ON/OFF ratio vs. $E_C$, and the ON/OFF ratio is the current ratio between LRS and corresponding shifted curve at the voltage of 7V in (a).

## 4. Conclusion.

The I-V curves in Fig. S3-S5 show that the HRS curve shifts further when encoded insulator capacitance $\beta$, remanent polarization $P_r$, or coercive field $E_c$ increases. And from both eq.2.2 and Fig. S3-5, we can find out the hyperbolic function in eq.2.2 limits the influence on the I-V curve shifting by the remanent polarization $P_r$ and the oxide capacitance $\beta$. On the other hand, we can find that the ON/OFF ratio increases drastically with the coercive field $E_c$ increasing, and the hyperbolic function does not limit it.

**Supplementary Note 2:** Linearization on the current-voltage relationship of ferroelectric diode

While Radu Berdan *et al.* proposed a logarithmic amplifier and a transimpedance amplifier by a well-fitted model in Ref. 3, we propose a simple min-max normalization and linear mapping procedure to encode the input voltage on software. With the $\log I$ - $V$ characteristics of the device shown in Fig. S6, the $\log I$ has excellent linearity with the voltage applied based on the nature of the conduction in a diode device [4], and the slopes of the plots in 16 different states of the device are consistent in the fitting.

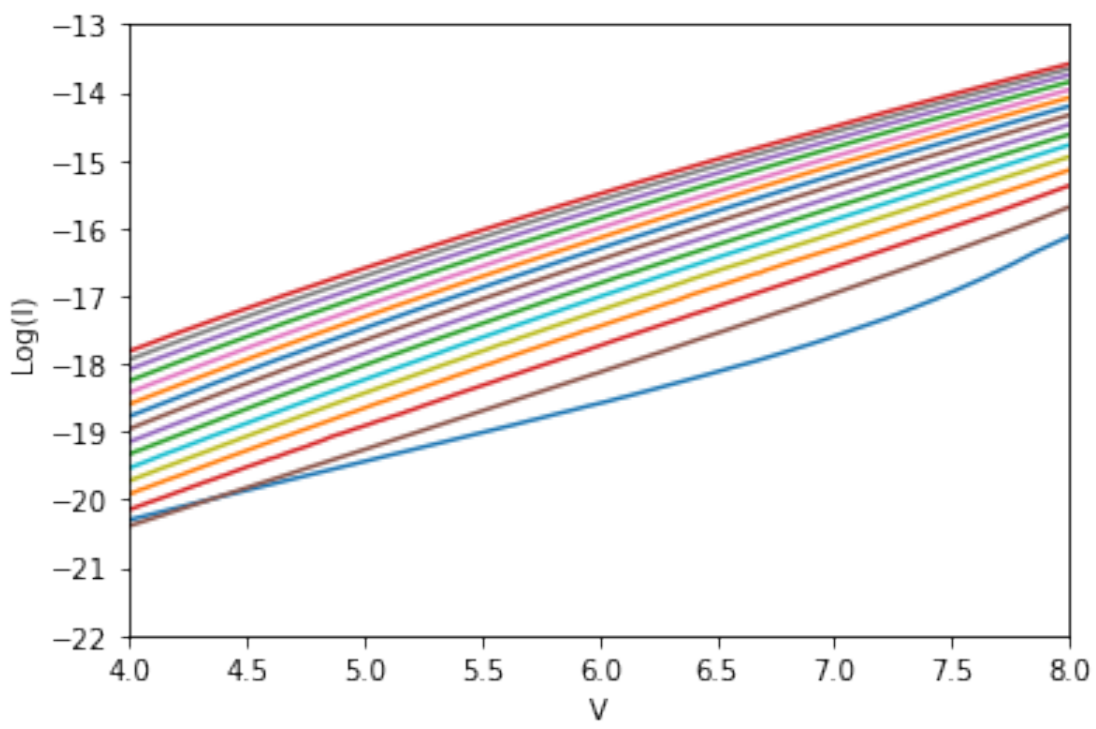

**Fig. S6.** Log(*I)* vs *V*. Current *I* measured by applying dc voltage sweeping to 16 different states.

Therefore, to linearize the $I$ - $V$ characteristics of a ferroelectric diode, we can have:

$$I = G_i \exp(\alpha V) \tag{3}$$

where the $G_i$ is a parameter that is related to the i'th conductance of the diode.

The constant slope $\alpha$ can be estimated through a linear regression method over the 16 distinct $\log I$ - $V$ characteristics of the ferroelectric diode device.

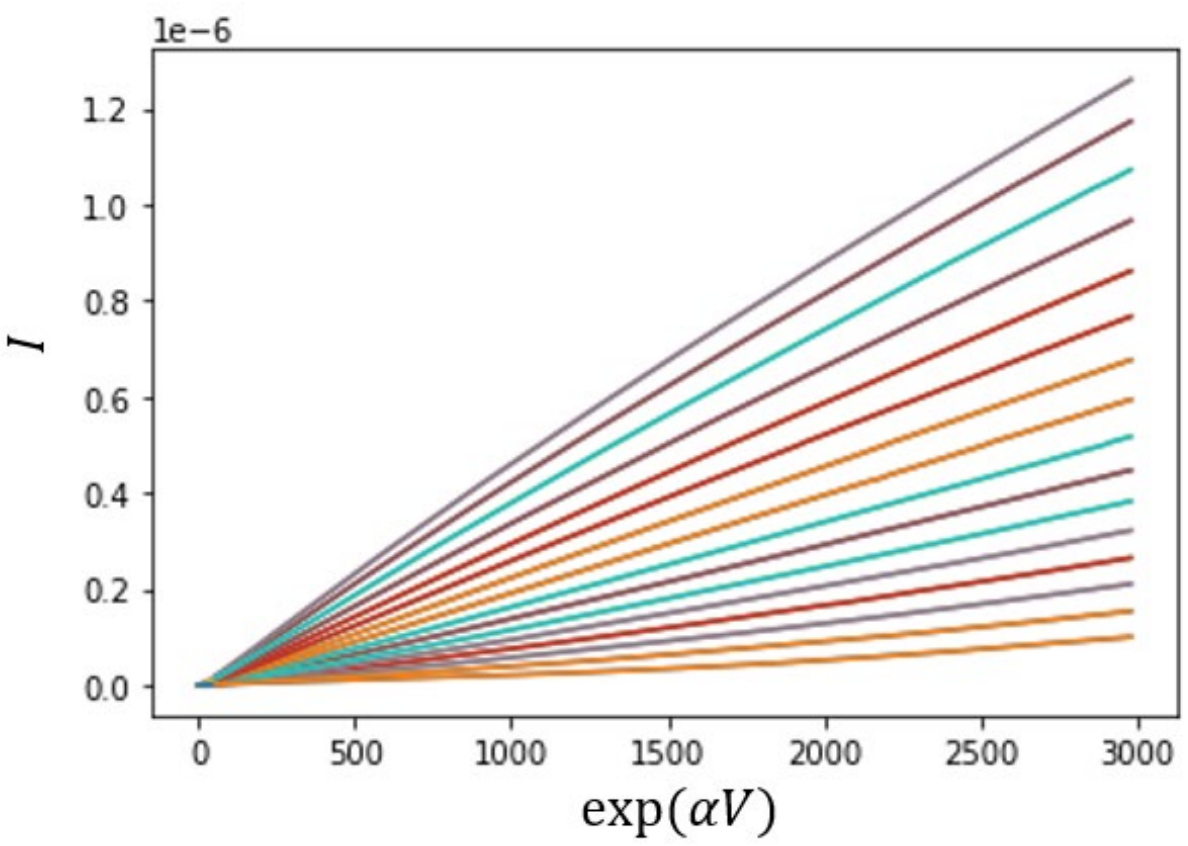

**Fig. S7.** Current *I* vs. $\exp(\alpha V)$ for 16 different states.

From Fig. S7, we can see that a ferroelectric diode's $I$ - $\exp(\alpha V)$ characteristics in 16 different states show superior linearity and intersection at the origin, which is the same as the $I$ - $V$ characteristics of an Ohmic resistor. The different slopes of all 16 states are proportional to the conductance of the 16 states, which also behave in a linear manner.

Given a input X from software in range [0, 1] and a read voltage window [$V_{min}$, $V_{max}$] on the FeD device, the min-max normalization and linear mapping procedure to encode the input voltage *f(X)* are described below:

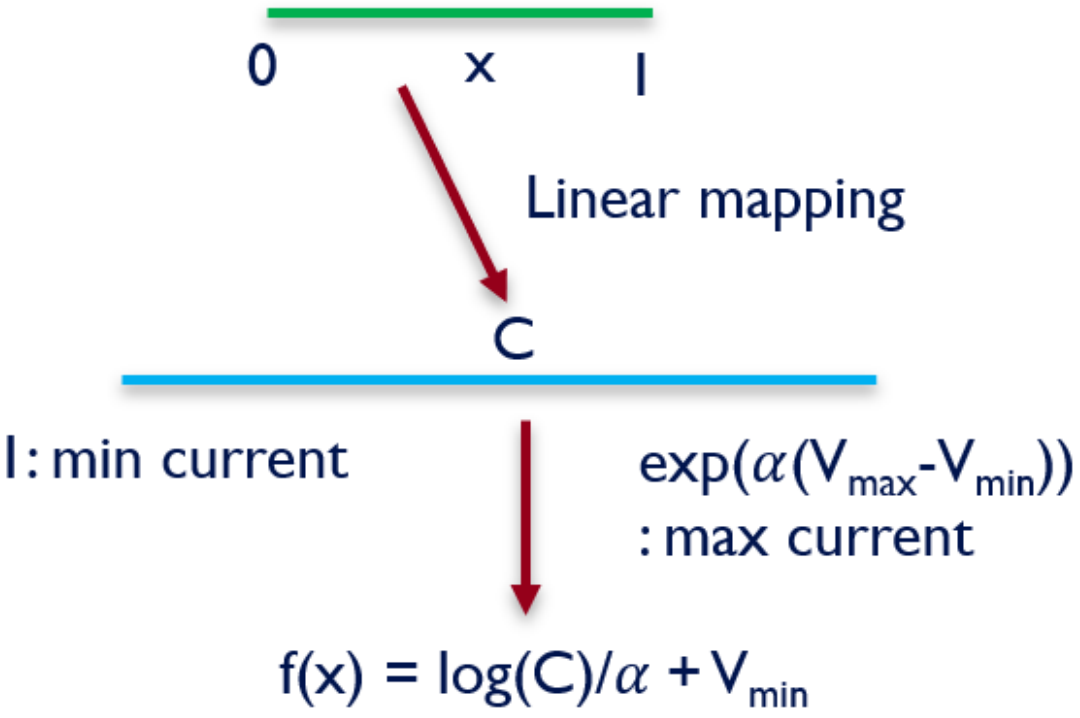

With the linear characteristics between current *I* and the function of voltage $f(X)$, we can linearly map the input to $f(X)$. For example, we have an input ranges from 0 to 1, we can map the input 1 to the $\max(\exp(\alpha V))$, input 0 to the $\min(\exp(\alpha V))$, and input between 0 and 1 to be distributed linearly between $\min(\exp(\alpha V))$ and $\max(\exp(\alpha V))$. Therefore, we can map each input to its corresponding $f(X)$, and decode $f(X)$ to find its corresponding Voltage $V$.

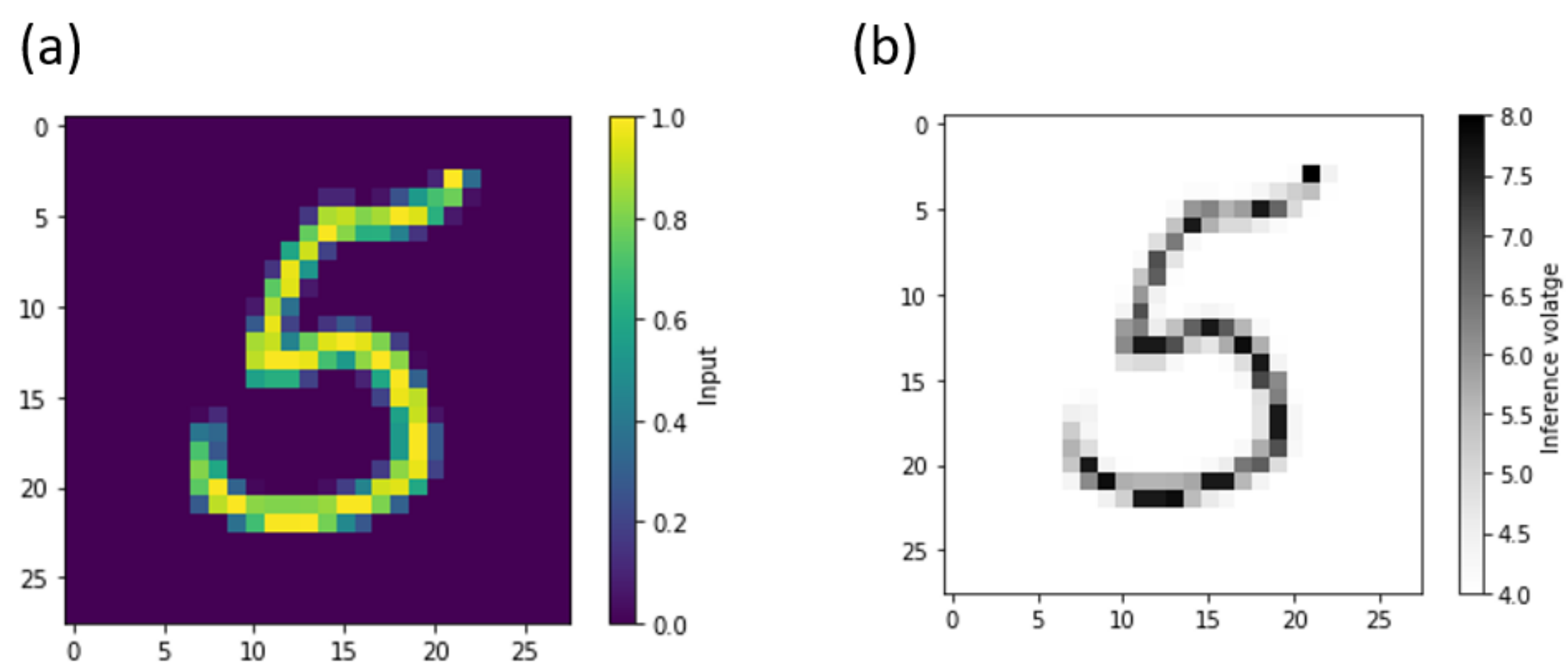

**Fig. S8.** (a) A picture of a number '5' as input from MNIST dataset. (b) Plotting after mapping and encoding the input signal into the realistic voltage $V$, the feature remains after converting the input signal to the voltage applied on the device.

Fig. S8(b) shows a transformation from the given input from the MNIST shown in Fig. S8(a), with the intensity as the input signal, to the encoded voltage amplitudes applied on the device. By this method, we could use the AlScN ferroelectric diode as a simple resistor in neural network computation by simply encoding the input $X$ to the function $V = f(X)$, then we could have the current $I$ measured as the output, shown below:

$$I = G_1 f(X_1) + G_2 f(X_2) + G_3 f(X_3) + \cdots \tag{4}$$

To verify the encoding scheme on the realistic device, we encode a series of inputs in the range of [0, 1] to our read voltage range [4V, 8V]. Then, we directly apply the encoded voltages on the ferrolectric diode devices and meaure the current as the output which is responding to the original input.

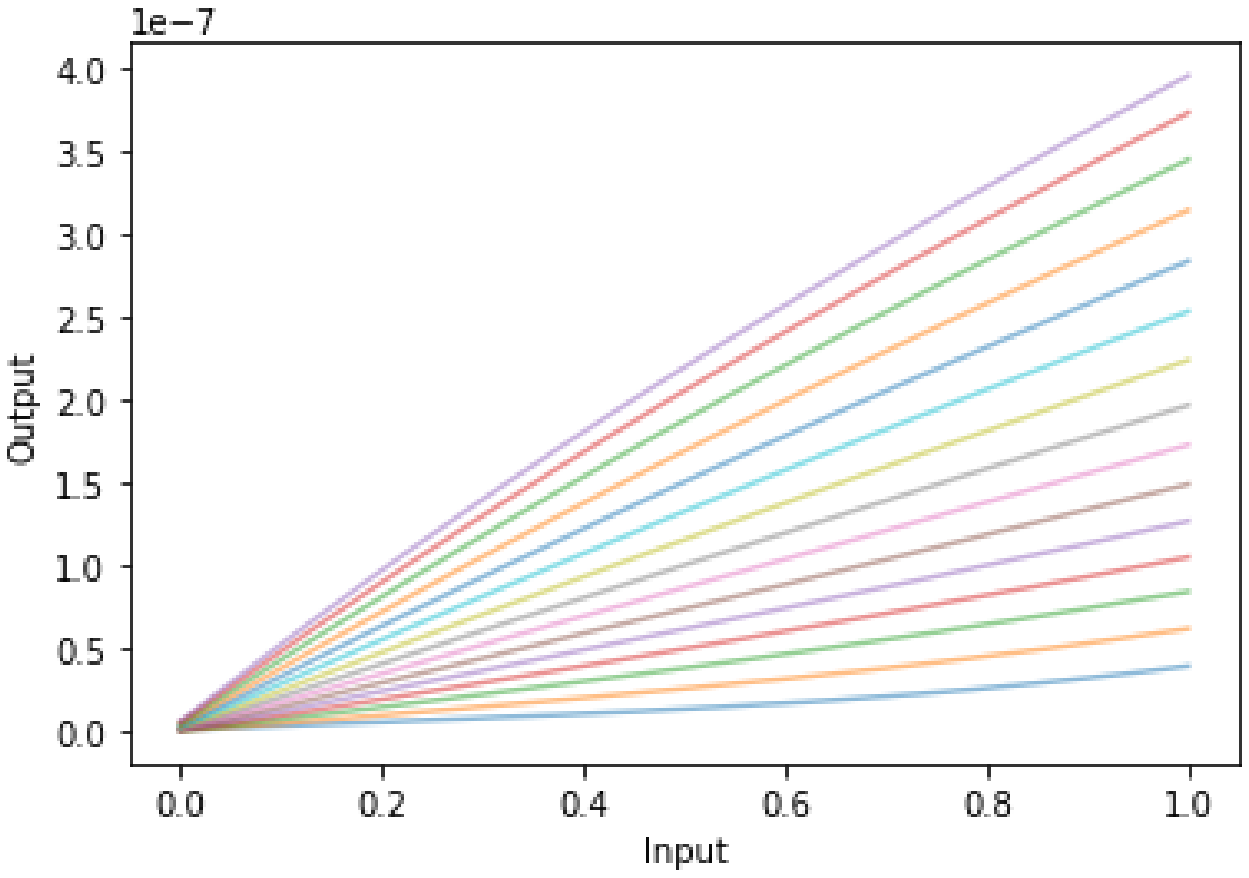

**Fig. S9.** The meaured current at the output with applying the encoded voltages on the ferrolectric diode devices which are responding to the original input.

As shown in Fig. S9, the output current presents a superior linearity on the input showing a $R^2$ score of 0.9998 for a linear fit, which verifies that we could use the AlScN ferroelectric diode as a simple resistor in neural network computation by simply using this linearization encoding method.

**Supplementary Note 3** Nonlinear weight update of ferroelectric diode

As we are typically using a linear quantization scheme when mapping the pre-trained weights to memristors, ideally, the amount of weight increase and weight decrease should be linearly proportional to the number of write pulses. However, the realistic devices reported in the literature do not follow such ideal trajectory where the conductance typically changes rapidly at the beginning then gradually saturates. This is one of the main reasons inhibiting highly accurate hardware matrix mulplication. Those non-linear weight updates of real devices can be evaluated by a factor A [5]:

$$G(N) = G_{\min} + (G_{\max} - G_{\min}) \frac{1-e^{-N/A}}{1-e^{-N_0/A}}$$

where $G_{min}$ and $G_{max}$ are the minimum and maximum conductance measured in the device, N is the pulse number we applied and $N_0$ is the maximum number we will apply. We could conclude from the above equaltions that as A decreases, the devices perform worse non-linear weight updates.

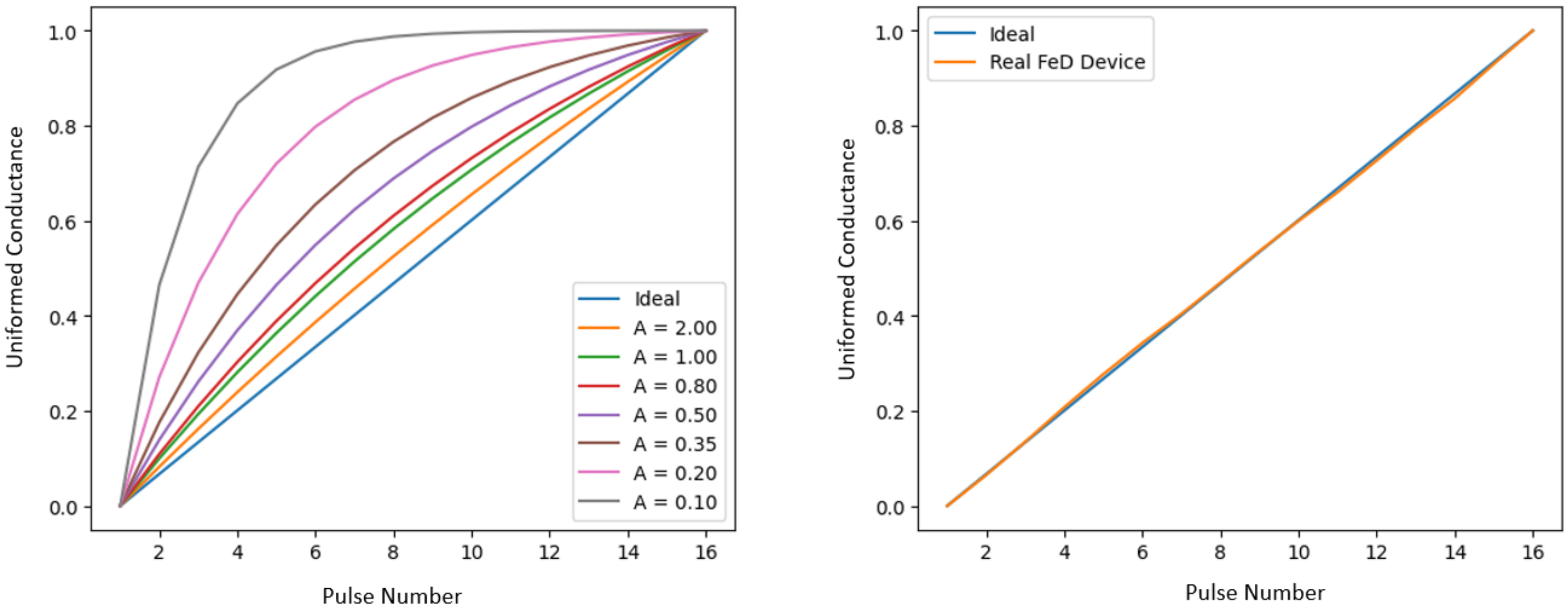

**Fig. S10.** (a) The normalized conductances by non-linear weight updates with different factor A. (b) The normalized conductances measured in the realistic FeD devices, which shows a near-ideal weight updates.

As shown in Fig. S10(b), the normalized conductances measured in our demonstrated FeD devices show near-ideal weight updates showing an A over 10 and $R^2$ score of 0.9997 to the ideal values.

| D1 | D2 | Store | SL | $\overline{SL}$ | Search | Results |
|---|---|---|---|---|---|---|
| negative-forward | negative-forward | Don't care | Vr | 0 | 1 | Match |
| negative-forward | negative-forward | Don't care | 0 | Vr | 0 | Match |
| positive-forward | negative-forward | 1 | Vr | 0 | 1 | Match |
| positive-forward | negative-forward | 1 | 0 | Vr | 0 | Mis |
| negative-forward | positive-forward | 0 | Vr | 0 | 1 | Mis |
| negative-forward | positive-forward | 0 | 0 | Vr | 0 | Match |

**Table. S1.** Voltage modes and encoding table for various values of stored states, search voltages, and search results for the demonstrated TCAM.

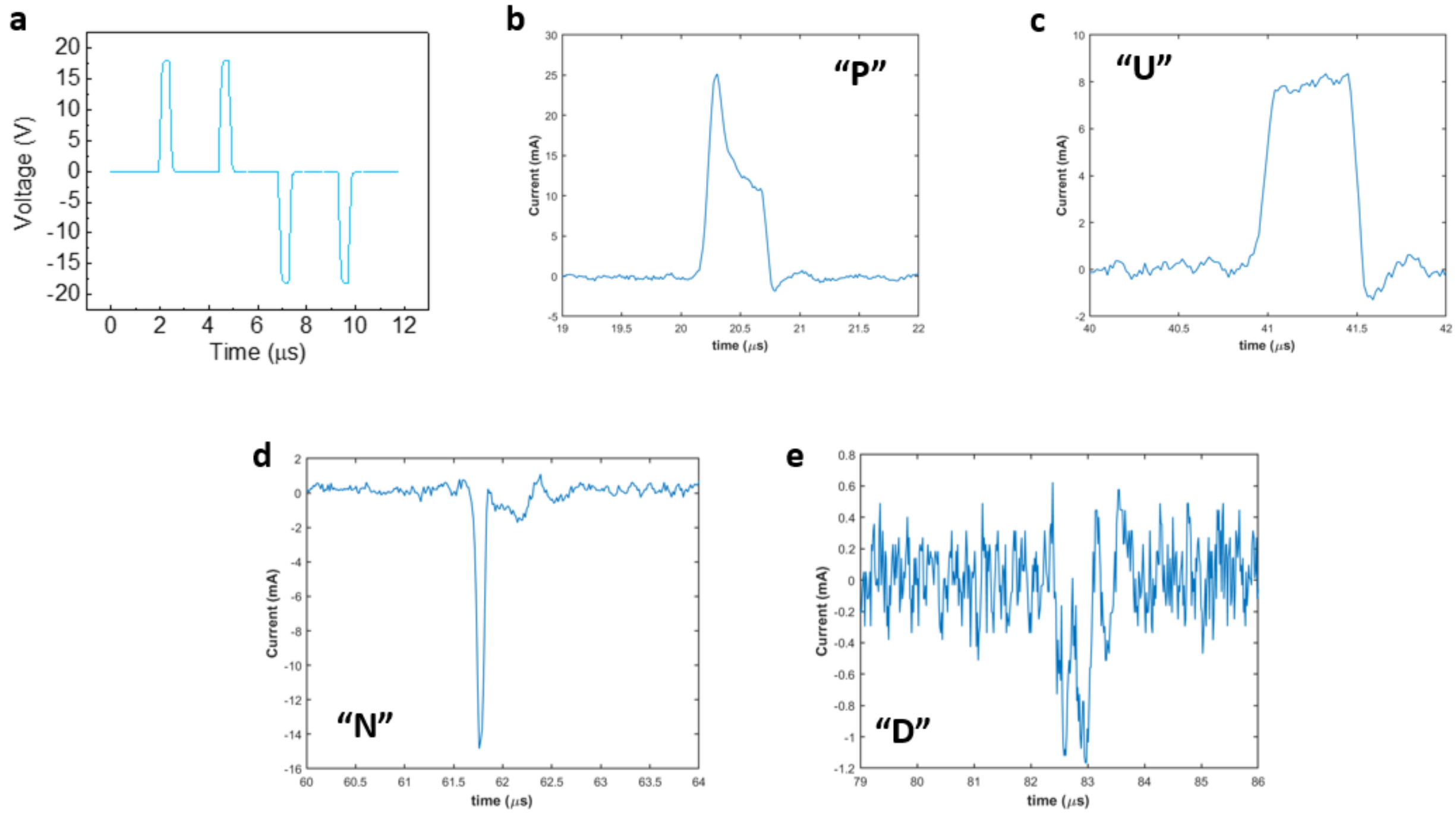

**Fig. S11.** a, Schematics of the signal sequences for the PUND measurements to differentiate the ferroelectric and non-ferroelectric contributions to the polarization. b, PUND current densities showing ferroelectric switching within 400 ns of the onset of the voltage switching pulse.

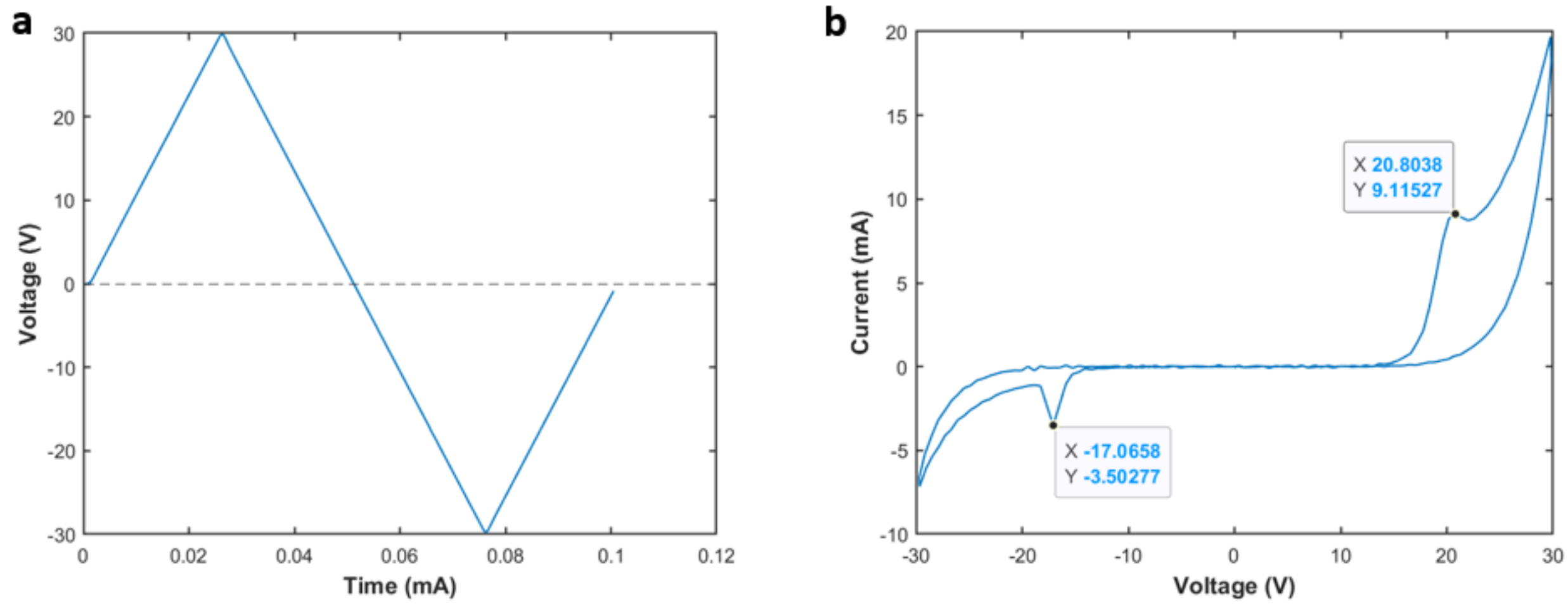

**Fig. S12. Dynamic current response in 45 nm AlScN**. a, Schematics of the signal sequences for the dynamic current response measurements to observe the ferroelectric switching induced current response. b, The current-voltage hysteresis loops (extracted from I-t curve) of a 45 nm thick AlScN corresponding to the signal sequences shown in a. The above plot shows a positive coercive field of +4.62 MV/cm and a negative coercive field of -3.79 MV/cm. Leakage optimization of the ferroelectric films is the subject of ongoing work.

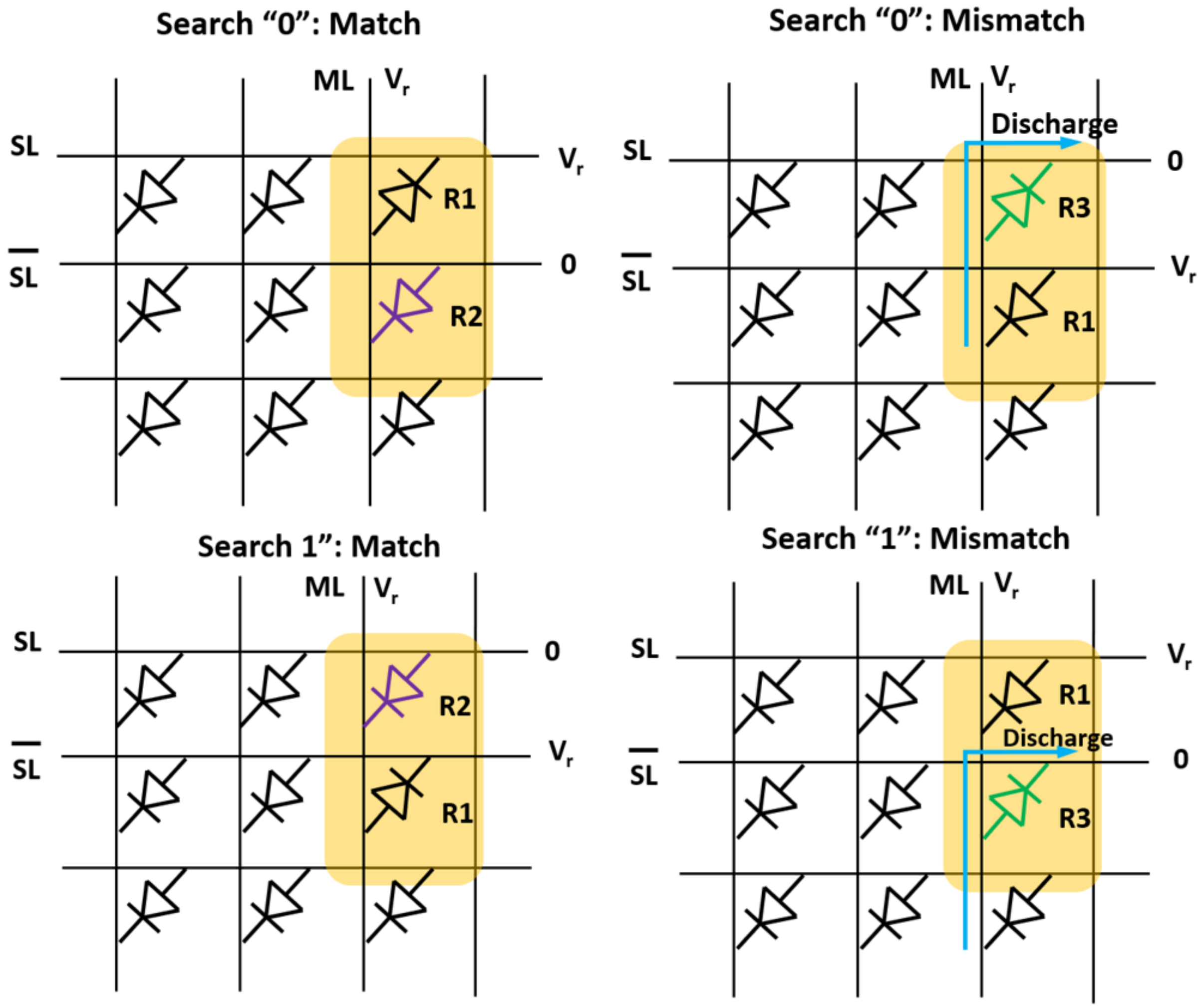

**Fig. S13.** The TCAM cell structure makes it natural to utilize the FeD crossbar memory array, in which the signal lines connecting to the anode and to the cathode are parallel in a bit search for the TCAM demonstration

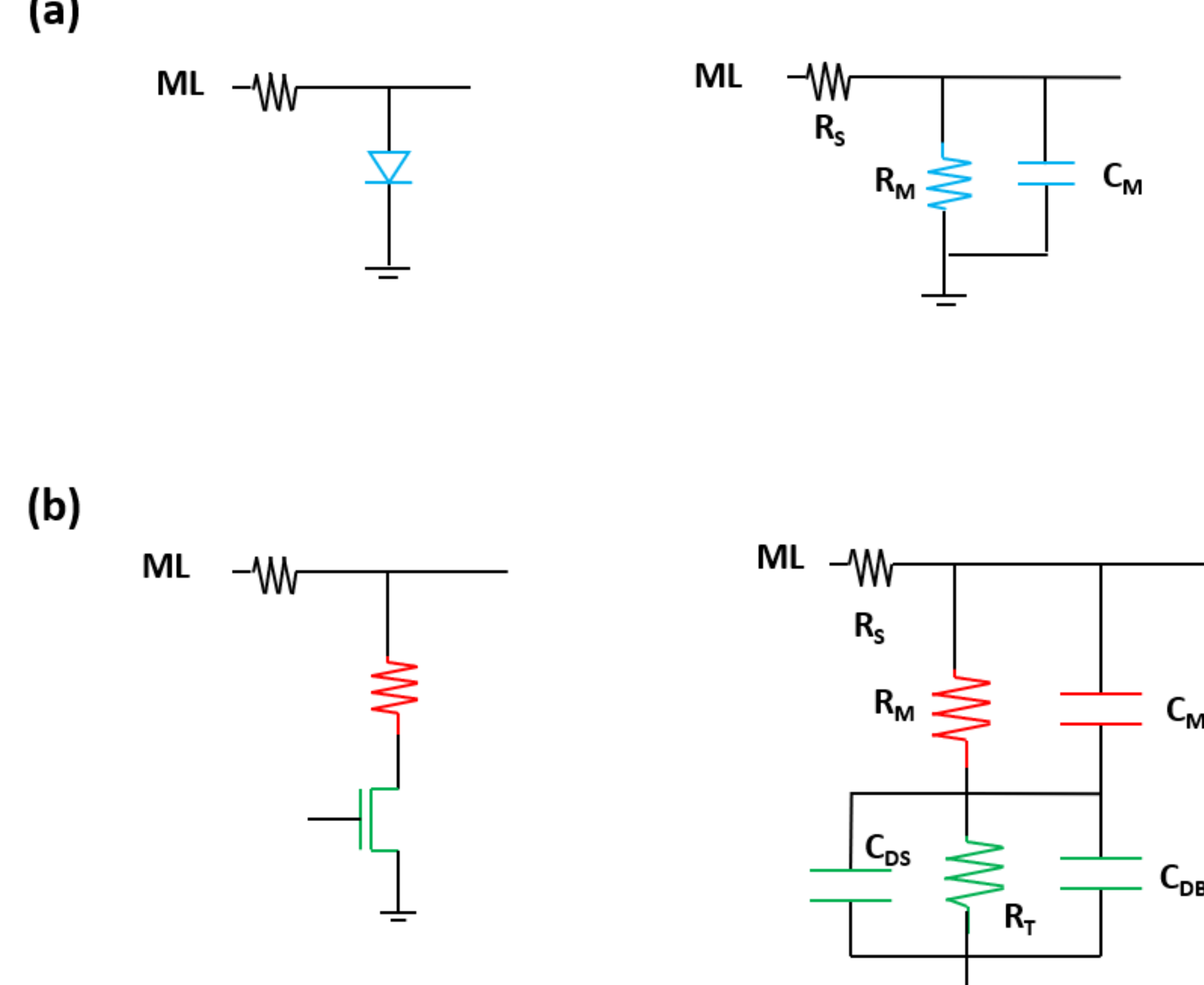

**Fig. S14.** The equivalent resistor-capacitor (RC) circuits have been shown in (a) for 2-FeD and in (b) for 2T-2R based TCAMs, which could explain why the search delay in our FeD-based TCAM can be reduced in comparison to prior TCAM architectures based on 2T-2R architecture.

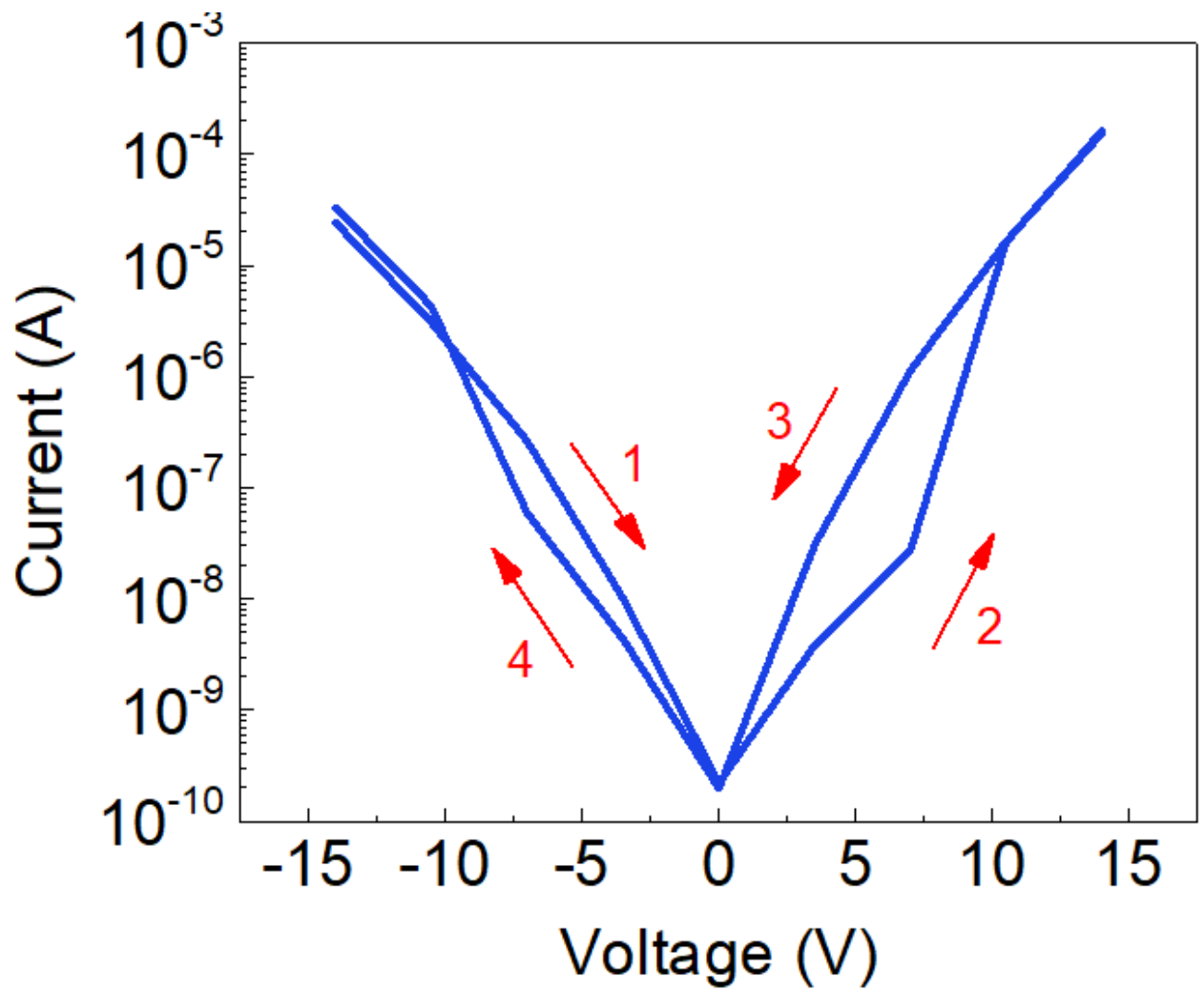

**Fig. S15. I-V curve of a single sweep over the demonstrated FeD device.** After being programmed by a positive voltage (sweep 1-2), the resistance changes from high-to-low, and the polarity of the device changes from a negative-forward diode to a positive-forward diode. Similarly, it can be observed in a negative voltage sweep that the polarity of the device changes from that of a positive forward diode to that of a negative-forward diode.

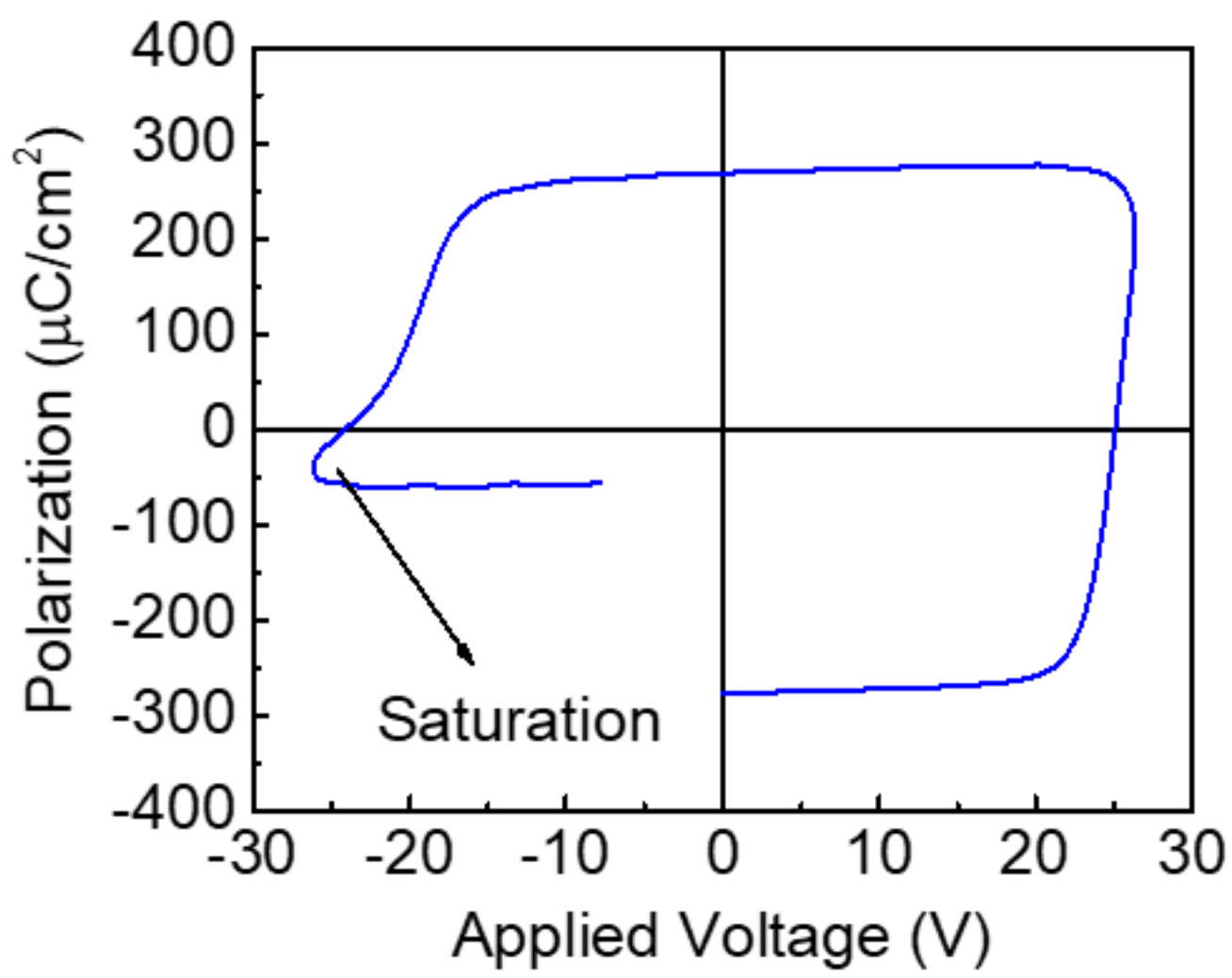

**Fig. S16. P-E loop measurements at 100 kHz in 45 nm AlScN.** A saturation in polarization is clearly observed at negative applied voltages while the leakage convolutes observation of saturation for positive voltages. The coercive field is observed to be slightly larger than the values observed in our DC measurements and the values reported in Ref. [6-8] at 300 K. We posit that this is because the coercive field is reported to significantly increase as frequency increases [6, 9-10].

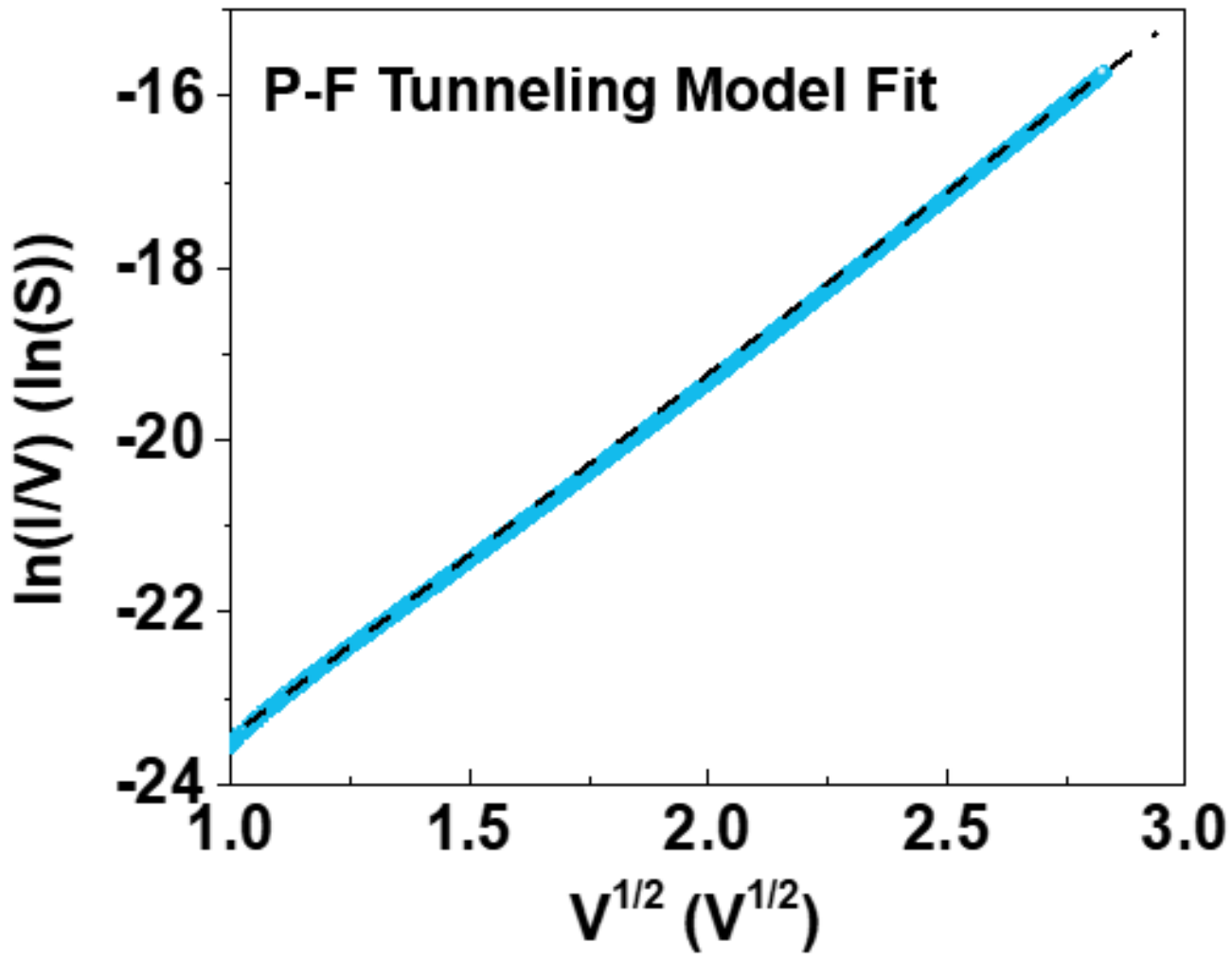

**Fig. S17.** Fitting of experimental current-voltage data to the Poole-Frenkel tunneling model showing a good fit.

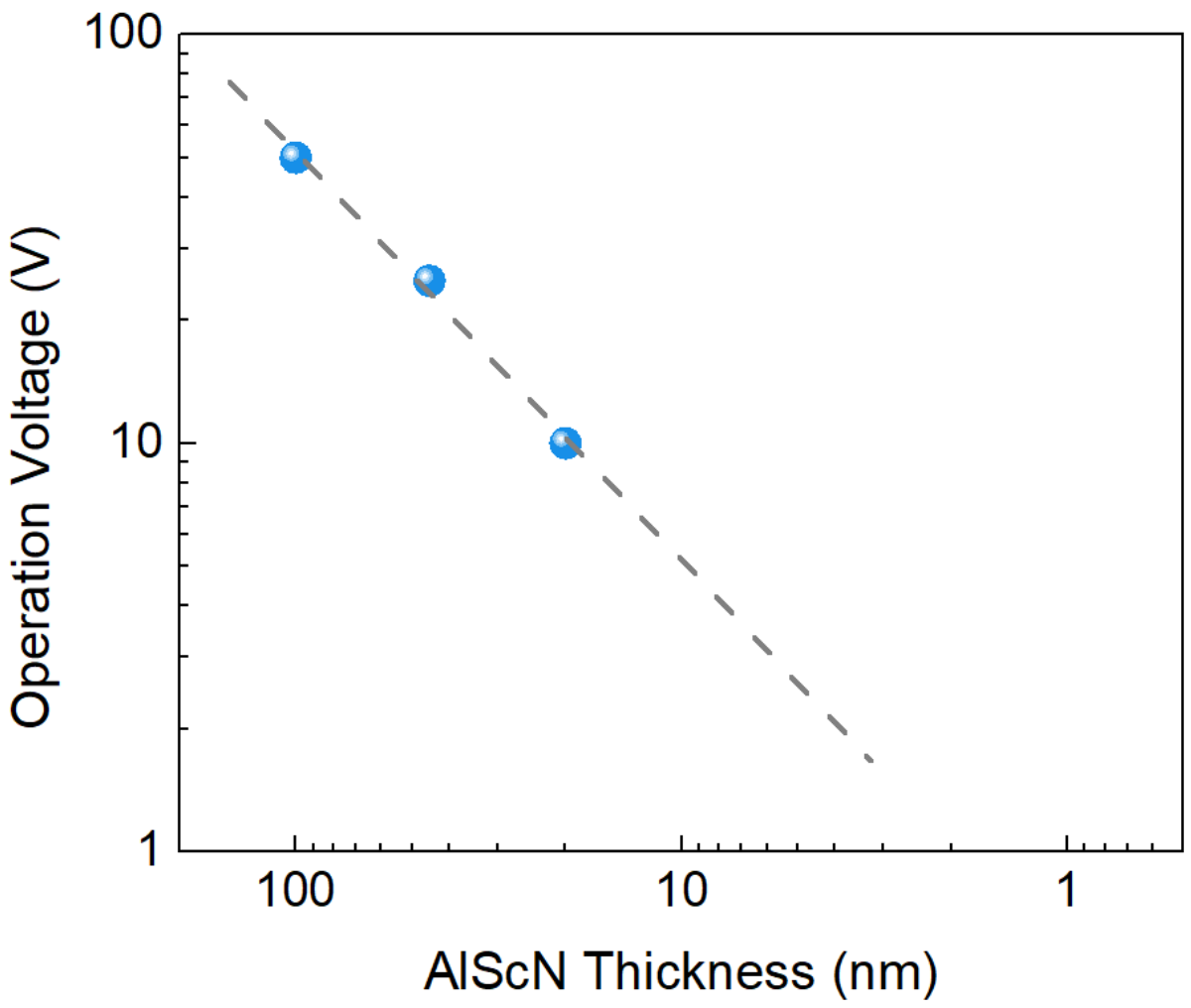

**Fig. S18.** Plot showing scaling down of the operating voltages of the FeD devices with reducing AlScN thickness.

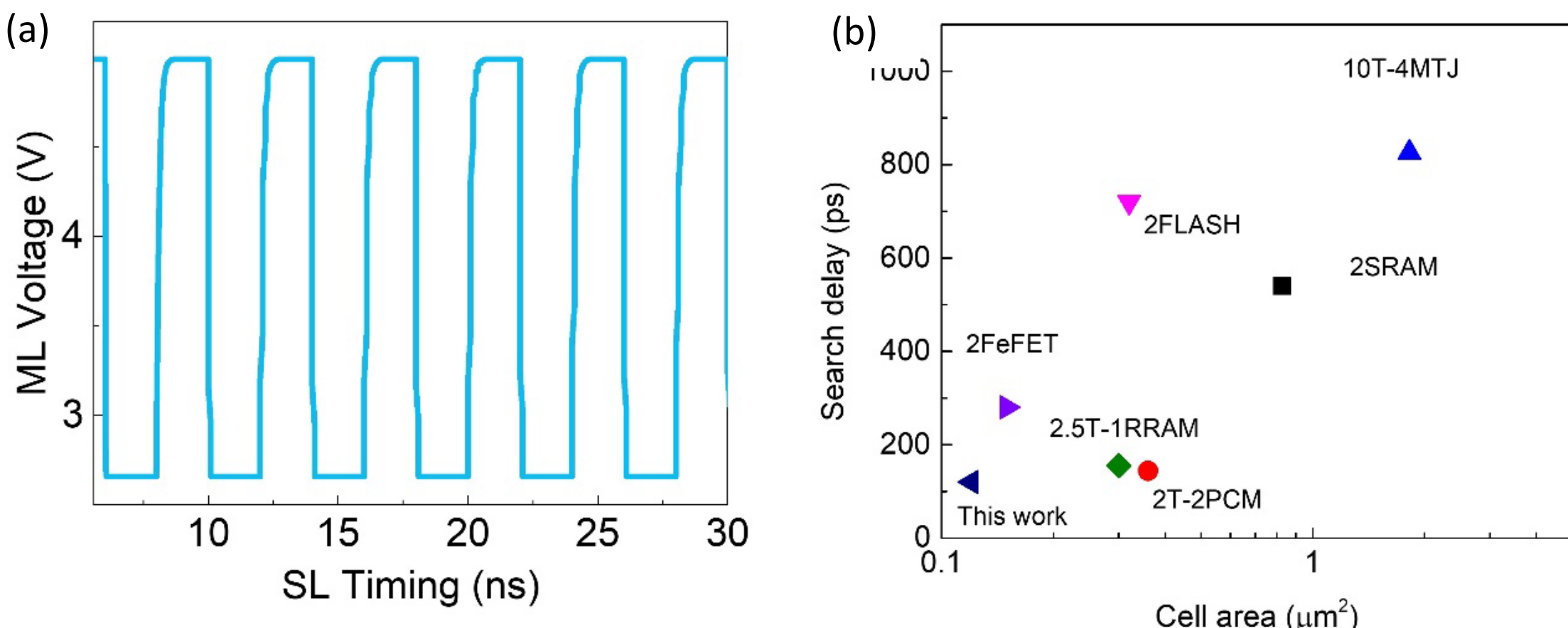

**Fig. S19.** (a) The transient SPICE simulation on match line voltage vs search line voltage timing. (b) Benchmark comparison chart of lateral footprint of various TCAM cells vs. search delay [11-15]: resistive random-access memories (RRAMs) [11], magnetic tunnel junction (MTJ) RAMs [12], floating gate transistor memory (FLASH) [13], phase change memories (PCMs) [14], and ferroelectric field-effect-transistor (FeFET) [15].